\documentstyle[prd,eqsecnum,aps,amsfonts,oldlfont]{revtex}
\begin{document}
\newcommand{\dal}{\square}
\newcommand{\CC}{{\Bbb C}}
\newcommand{\MM}{{\Bbb M}}
\newcommand{\QQ}{{\Bbb Q}}
\newcommand{\RR}{{\Bbb R}}
\newcommand{\TT}{{\Bbb T}}
\newcommand{\VV}{{\Bbb V}}

\newcommand{\cA}{{\cal A}}
\newcommand{\cC}{{\cal C}}
\newcommand{\cF}{{\cal F}}
\newcommand{\cL}{{\cal L}}
\newcommand{\cM}{{\cal M}}
\newcommand{\cO}{{\cal O}}
\newcommand{\cS}{{\cal S}}
\newcommand{\eps}{\epsilon}
\newcommand{\chib}{{\bar\chi}}
\newcommand{\epsb}{{\bar\epsilon}}
\newcommand{\mub}{{\bar\mu}}
\newcommand{\psib}{{\bar\psi}}
\newcommand{\pib}{{\bar\pi}}
\newcommand{\taub}{{\bar\tau}}
\newcommand{\del}{\partial}
\newcommand{\tfrac}[2]{{\textstyle{{#1}\over{#2}}}}
\newcommand{\half}{\tfrac{1}{2}}
\newcommand{\quart}{\tfrac{1}{4}}
\newcommand{\qed}{\relax}
\newcommand{\spinor}[1]{#1^{AB\ldots C}_{B'\ldots C'}}
\newcommand{\norm}[1]{||#1||}
\newcommand{\Khat}{\widehat{K}}
\newcommand{\Lhat}{\widehat{L}}
\newtheorem{prop}{Proposition}[section]
\newtheorem{lem}[prop]{Lemma}
\newtheorem{thm}[prop]{Theorem}
\newtheorem{cor}[prop]{Corollary}
\newtheorem{defn}[prop]{Definition}
\renewcommand{\Im}{\mathop{\rm Im}\nolimits}
\title{
On a class of consistent linear higher spin equations
on curved manifolds}
\author{J\"org Frauendiener}
\address{
Max-Planck-Institut f\"ur Gravitationsphysik,
Albert-Einstein-Institut,
Schlaatzweg 1,
D--14473 Potsdam,
Germany}
\author{George A. J. Sparling}
\address{
Department of Mathematics and Statistics,
University of Pittsburgh,
Pittsburgh, PA 15260,
USA}
\maketitle

\begin{abstract}
We analyze a class of linear wave equations for odd half spin that
have a well posed initial value problem. We demonstrate consistency of
the equations in curved space-times. They generalize the Weyl neutrino
equation. We show that there exists an associated invariant exact set
of spinor fields indicating that the characteristic initial value
problem on a null cone is formally solvable, even for the system
coupled to general relativity. We derive the general analytic solution
in flat space by means of Fourier transforms. Finally, we present a
twistor contour integral description for the general analytic solution
and assemble a representation of the group $O(4,4)$ on the solution
space.
\end{abstract}

\section{Introduction}

It is a well known fact that many spinor equations that are perfectly
well behaved in (flat) Minkowski space can not be translated to a
general four-dimensional curved background manifold. This happens
e.g., for the zero rest-mass (zrm) fields with spin $s>1$ and for the
twistor equation. In these cases the appearance of the Buchdahl
conditions (\cite{Buchdahl-1958}, \cite{PenroseRindlerI}) imposes
algebraic conditions relating any solution of the field equations to
the (conformal) curvature of the manifold. In the zrm case, these
conditions are very restrictive in that they limit the solution space
of the equations for a general given background manifold. In the case of
the twistor equation one finds that solutions can exist only on
algebraically special manifolds of type $N$ or $O$. Recently
\cite{Penrose-1991}, there has been some interest in the case
$s=3/2$, the Rarita-Schwinger system, for which the consistency
condition is just Ricci-flatness of the manifold such that one can
regard the vacuum Einstein equations as integrability conditions for
this system of spinor equations.

Despite this very interesting approach to the vacuum Einstein
equations and its relation to twistor theory we want to present here a
class of spinor equations which does not have the drawback of being
well defined only on flat space-time (cf. also \cite{sparling-1992}). In
particular, these are linear equations for a spinor field of half
integer spin $s$, which include the Weyl neutrino equation as the case
$s=(1/2)$.  The general equation we shall consider may be written as follows:
\begin{equation}
\del^{A(A'}\phi_{AB\ldots CD}^{B'\ldots C'D')} = 0.
\label{feq}
\end{equation}

Here the operator $\del_{AA'}$ is the standard Levi-Civit\`a spin
connection and the spinor field $\phi_{AB\ldots CD}^{B'\ldots C'D'}$ is
totally symmetric and has $m+1$ unprimed indices and $n$ primed
indices, with $m$ and $n$ non-negative integers.  A simple count shows
that the field $\phi_{AB\ldots CD}^{B'\ldots C'D'} $ has $(m + 2)(n + 1)$
components at every point, whereas the number of field equations is
$(m + 1)(n + 2)$.  So the excess number of field equations vis \`a vis
components is $m - n$.  In the case $m < n$, we have fewer field
equations than components.  Typically this leads to gauge freedom (at
least in flat space): for example in the case $(m, n) = (0, 1)$,
equation (\ref{feq}) gives the self-dual Maxwell equations for a potential
$\phi_a$ which has the gauge freedom $\phi_a \mapsto \phi_a +
\del_a \phi$ with $\phi$ an arbitrary scalar field.  Next in the
case $m > n$, we have an over determined system.  This leads to
integrability conditions and possible inconsistencies in a general
curved spacetime.  The classic example is the case $n = 0$, in which
case equation (\ref{feq}) becomes just the standard zrm for spin $(m +
1)/2$, which is inconsistent in general, at least for $m > 4$.  The
subject of this work is the case $m = n$, where there are exactly as
many equations as there are field components, so one might expect that
there are no non-progagating degrees of freedom and no constraints.
Indeed we will show that these equations possess the following
properties:

\begin{itemize}
\item[--] They can be derived from a variational principle;
\item[--] They are conformally invariant;
\item[--] The Cauchy problem is well posed in flat and in an arbitrary
curved space;
\item[--] There exists an equivalent exact set of spinor fields
\cite{Penrose-1980} for the fields when propagating on a curved
background and also when they are coupled to
gravity, which means that the characteristic initial value problem is
formally well posed;
\item[--] The fields propagate along null hypersurfaces in flat space.
However, they do not satisfy the wave equation $\dal\phi=0$ but an
equation $\dal^{m+1}\phi=0$ instead, where $\dal$ is the d'Alembert
operator corresponding to the spin connection;
\item[--] The general solution in Minkowski space can be
given using a variation of Fourier transforms;
\item[--] A twistor description for analytic flat space solutions can
be given.
\end{itemize}
The plan of the paper is as follows: in section 2. we present the
variational principle and derive the energy momentum tensor. To do
this we need to explain and extend a formalism given elsewhere
\cite{jf-1992-1}. In section 3. we discuss the Cauchy
problem and show the existence
of solutions. In section 4. we explain the relation with exact sets and
discuss the characteristic initial value problem. Section 5. is
devoted to the general solution in Minkowski space and in section 6.
we show how to obtain solutions by performing contour integrals in
twistor space thereby establishing an isomorphism between analytic
flat space solutions and certain cohomology groups for suitable domains in
projective twistor space.

\section{The variational principle and the energy momentum tensor}

In this section we will derive the spinor equations from a variational
point of view. We will use a formalism described in \cite{jf-1992-1}
to derive our results. In order to introduce the
notation and also to transcribe the various formulae to apply to
spinors we will briefly review its basic features.

The starting point is an $SO(1,3)$ principal bundle $O(\cM)$ over
space-time $\cM$
carrying a tensorial one-form $\theta^a$ with values in $\RR^4$ and a
connection form $\theta^a{}_b$ with values in the Lie algebra
$so(1,3)$ of the structure group. There also exists a constant matrix
$\eta_{ab}$ of signature $(1,3)$ that is used to construct the Lorentz
metric on $\cM$. We will require that the connection be torsion free,
i.e., $D\theta^a=0$; $D$ is the exterior covariant derivative. Making
use of the standard $2{\rm -to-}1$ epimorphism of $SL(2,\CC)$ onto the
orthochronous Lorentz group to enlarge the structure group of the
bundle and employing abstract index notation we may write $\theta^a=
\theta^{AA'}$, $\theta^a{}_b=-\eps^A{}_B \theta^{A'}{}_{B'} -
\eps^{A'}{}_{B'} \theta^{A}{}_{B}$, thus defining the unprimed and
primed spin connections, symmetric in their respective indices (the
sign is chosen in order to conform with other references). Then the torsion
free condition is (note, that we suppress the wedge because it will be
the only product we use between forms)
\begin{equation}
0=d\theta^{AA'} + \theta^A{}_B \theta^{BA'} + \theta^{A'}{}_{B'}
\theta^{AB'}.\label{2.1x}
\end{equation}
We introduce a set of vector fields $\del_{AA'}$, $\del_A{}^B$ dual to
the forms  $\theta^{AA'}$ and $\theta^A{}_B$. Their action will be
extended from functions to indexed quantities by requiring that they
be derivations of the algebra of indexed forms annihilating
$\theta^{AA'}$ and $\theta^A{}_B$, see \cite{jf-1993-2} for further
details.

We define a variation at $\theta^a$ and $\theta^a{}_b$ as the derivative
at $\lambda=0$ of one-parameter families $\theta^a(\lambda)$ and
$\theta^a{}_b(\lambda)$ with $\theta^a(0)=\theta^a$ and
$\theta^a{}_b(0)=\theta^a{}_b$.  Denoting the variations by $\chi^a$ and
$\chi^a{}_b$ we find that $\chi^a$ is a tensorial one-form with values
in $\RR^4$, $\chi^a{}_b$ is a tensorial one-form with values in
$so(1,3)$. Using spinor indices and decomposing into irreducible parts
we have
\begin{equation}
\chi^{AA'}=\sigma^A{}_B{}^{A'}{}_{B'} \theta^{BB'} + \quart \sigma
\theta^{AA'} + \tau^A{}_B \theta^{AB'} + \taub^{A'}{}_{B'}
\theta^{BA'} .\label{2.2x}
\end{equation}
The variations are not independent; in fact, the torsion free
condition entirely fixes the variation $\chi^A{}_B=\chi^A{}_{BCC'}
\theta^{CC'}$ of the connection for given $\chi^a$:
\begin{equation}
\chi_{ABCC'}= -\del^{E'}_{(A} \sigma_{B)CE'C'} + \quart \eps_{C(B}
\del_{A)C'} \sigma + \del_{CC'} \tau_{AB} .\label{2.3x}
\end{equation}
Spinor fields can now be considered as tensorial functions with
values in the appropriate representation of $SL(2,\CC)$. In particular
we consider totally symmetric fields $\psi^{AB\ldots C}_{B'\ldots C'}$
with $m$ primed
and $m+1$ unprimed indices. Note however, that we will switch back and
forth between viewing the fields as tensorial functions on the bundle
and as spinor fields on space-time $\cM$ (section in an appropriate
associated bundle). With these preparations we can now write
down the following horizontal four-form
\begin{equation}
\cL\equiv \Im\left(\psib^{A'B'\ldots C'}_{B\ldots C} D\psi^{AB\ldots
C}_{B'\ldots C'} \right) \Sigma_{AA'},\label{2.4x}
\end{equation}
with $\Sigma_a=(1/6)\eps_{abcd}\theta^b\theta^c\theta^d$. Note that we
consider our spinors to be commuting quantities. If instead we would
need to have anticommuting spinors we would use the real part instead
of the imaginary part in equation (\ref{2.4x}).  Since $\cL$
can be considered as the pullback of a unique globally defined four-form on
$\cM$ we may define the action $\cA$ as the integral of $\cL$ over $\cM$:
$\cA\equiv \int_\cM\cL$. It is then easily verified that the variations
of $\cA$ with respect to $\psib$ ($\psi$) give the equation (and its
complex conjugate):
\begin{equation}
D\psi^{AB\ldots C}_{(B'\ldots C'}\Sigma_{A')A}=0. \label{2.5x}
\end{equation}
Now $D=\theta^a\del_a$ and $\theta^a\Sigma_b = \delta^a_b \Sigma$ with
$\Sigma=(1/24)\eps_{abcd}\theta^a\theta^b\theta^c\theta^d$. Using this
and stripping off the form $\Sigma$, we obtain the desired field
equation
\begin{equation}
\del_{A(A'}\psi^{AB\ldots C}_{B'\ldots C')}=0. \label{feqn}
\end{equation}
Let us now discuss some of the basic properties of this field
equation. First, we note that for $m=0$ this is just the Weyl neutrino
equation $\del_{AA'}\nu^A=0$, which is the zrm field equation for spin
$(1/2)$. Just as the neutrino equation the general equations are
conformally invariant if we assume a transformation of the fields with
conformal weight $-2$. We define the other irreducible parts of the
covariant derivative of $\psi$,
\begin{eqnarray}
\lambda^{AB\ldots CD}_{B'\dots C'D'} &=
\del^{(D}_{(D'}\psi^{AB\ldots C)}_{B'\ldots C')}, &\label{lambdadef} \\
\mu^{AB\ldots C}_{B'\ldots C'} &=
\del^{E'(A}\psi^{B\ldots C)}_{E'B'\ldots C'}, & \label{mudef} \\
\nu^{AB\ldots C}_{B'\ldots C'} &=
\del^{E'}_{E}\psi^{EAB\ldots C}_{E'A'\ldots C'}. &\label{nudef} \\
\end{eqnarray}
Then we have the expansion
\begin{equation}
\del^E_{E'}\psi^{AB\ldots C}_{B'\ldots C'}=
\lambda^{EAB\ldots C}_{E'B'\ldots C'} - {m\over m+1} \eps_{E'(B'}
\mu^{EAB\ldots C}_{\ldots C')} - {m\over m+2} \eps_{E'(B'} \eps^{E(A}
\nu^{B\ldots C)}_{\ldots C')}.\label{expansion}
\end{equation}
By virtue of the field equation these fields satisfy the following
relations (among others):
\begin{eqnarray}
{m\over m+1}\del_{A(A'}\nu^{AC\ldots D}_{C'\ldots D')}&=&
\dal_{AB}\psi^{ABC\ldots D}_{A'C'\ldots D'}, \label{Mbnu} \\
{m\over m+1}\del_{(A'}^{(A}\nu^{BC\ldots D)}_{C'\ldots D')}&=&
\dal^{(A}{}_{E}\psi^{BC\ldots D)E}_{A'C'\ldots D'}
-\half\dal\psi^{ABC\ldots D}_{A'C'\ldots D'}, \label{Lnu} \\
\del_{A(A'}\lambda^{AB\ldots CD}_{B'\ldots C'D')} &=&
\dal_{(A'B'}\psi^{BC\ldots D}_{C'\ldots D')},\label{Mblambda} \\
\del_{A}^{A'}\lambda^{AB\ldots CD}_{A'C'\ldots D'} &=&
-{1\over 2(m+1)}\dal\psi^{BC\ldots D}_{C'\ldots D'} -
\dal_A{}^{(B}\psi^{C\ldots D)A}_{C'\ldots D'} +  \\
&& {m\over m+1}\del_{(C'}^{(B}\psi^{C\ldots D)}_{\ldots D')}
+ {m\over m+1} \dal^{E'}_{(C'}\psi^{BC\ldots D}_{\ldots D')E'}, \\
\del_{A(A'}\mu^{AB\ldots CD}_{B'\ldots C')} &=&
{1\over 2(m+2)}\dal\psi^{B\ldots CD}_{A'B'\ldots C'} +
\dal_{(A'}^{D'}\psi^{BC\ldots D}_{B'\ldots C')D'} + \\
&& {{m+1}\over m+2} \dal_{A(B}\psi^{C\ldots D)A}_{A'B'\ldots C'} +
{1\over m+2} \del^{(B}_{(A'}\nu^{C\ldots D)}_{B'\ldots C')}, \label{Mbmu}
 \\
\del_{A}^{B'}\mu^{ABC\ldots D}_{B'C'\ldots D'} &=&
-{{m+1}\over m+2}\left\{ \del^{B'(B}\nu^{C\ldots D)}_{B'C'\ldots D'}
+ \dal^{C'B'}\psi^{BC\ldots D}_{B'C'\ldots D'} \right\}.
\end{eqnarray}

In these formulae we have used the spinor curvature derivations
$\dal_{AB}$ and $\dal_{A'B'}$ as
defined in \cite{PenroseRindlerI}. To further analyze the situation
it is very convenient to introduce four differential
operators $L$, $M$, $M'$ and $N$ acting on irreducible spinor fields
by taking the covariant derivative and then projecting onto one of the
four possible irreducible parts (see \cite{jf-1993-2}
for a rigorous definition and further details). Thus, for a field
$\phi$ with $p$ unprimed and $p'$ primed indices and all its indices
down, we identify $L\phi\doteq \del_{(E'(E} \psi_{A\ldots B)C'\ldots D')}$,
$M'\phi\doteq -p' \del^{E'}_{(E}\psi_{A\ldots B)C'\ldots E'}$,
$M\phi\doteq -p
\del^{E}_{(E'}\psi_{C'\ldots D')A\ldots E}$, $N\phi\doteq pp'
\del^{EE'}\psi_{A\ldots EC'\ldots E'}$. These operators obey certain
commutation rules, most of which are trivial in flat space. The
nontrivial ones are
$\left[L,N\right]\phi = -\half(p+p'+2)\dal\phi$ and $\left[ M,M'
\right]\phi = -\half (p-p')\dal\phi$. There is a further relation:
$LN\phi - MM'\phi -\half p(p'+1)\dal\phi =0$. The wave operator commutes
with all the derivative operators. In flat space, the equations
(\ref{Mblambda}), (\ref{Mbnu}), (\ref{Mbmu}) and (\ref{Lnu}) above
can be written as follows:
\begin{equation}
M'\lambda=0,\quad M'\nu=0,\quad M'\mu = \half \dal\psi,
\quad N\lambda=-m(m+1)L\nu +\half (2m+3) \dal\psi,
\quad L\nu= -\frac{m+1}{2m} \dal\psi. \label{LMN}
\end{equation}
We observe that $\nu$ and $\lambda$ are spinor fields of the same class
as $\psi$ obeying the same equation. In contrast to the zrm case, the
field $\psi$ does not obey the wave equation $\dal\psi=0$ (unless
$m=0$, because then $\mu=0$). However,
with these preparations it is now easy to prove the
\begin{prop}\label{prop2.1}
 Given a smooth spinor field $\psi$ with $m$ primed
and $m+1$ unprimed indices subject to the field equation (\ref{feq}),
then $\dal^{m+1}\psi=0$ in flat space.
\end{prop}
\noindent{\it Proof:\/} We prove this by induction on $m$. The case
$m=0$ is the Weyl neutrino equation for which the assertion is true.
Now assume it is true for $(m-1)$, then $\dal^m\nu=0$. But then $0=
L\dal^m\nu= \dal^m LN \psi = \dal^m (\half(m+1)^2\dal)\psi = \half
(m+1)^2 \dal^{m+1}\psi$. \qed

Finally, we want to derive the energy momentum tensor of these spinor
fields. This is usually done by considering the action as depending on
the metric of the background manifold and then varying with respect to
that metric. The result is the natural object that would appear on the
right hand side of the Einstein equations if the system were coupled to
gravity. In our case we can not regard the action as depending on the
metric, we have to take it depending on the canonical one-form. Then
the variation of the action with respect to $\theta^a$ contains terms
proportional to $\sigma_{ABA'B'}$, $\tau_{AB}$ and $\sigma$. The
functional derivative of the action with respect to $\sigma_{ABA'B'}$
is the trace free part of the energy momentum tensor while taking the
functional derivative with respect to $\sigma$ gives the trace part.
However, in the present case, we expect this term to vanish due to the
conformal invariance of the equation and the functional
derivative with respect to $\tau_{AB}$ will be seen to vanish as well.
This is related to the fact that the connection is required to be
torsion free. The variation of the action with respect to
$\theta^{AA'}$ is $\delta\cA=\Im \int \delta\cL$ with
\begin{equation}
\delta\cL= \left\{
(m+1)\psib^{A'B'\ldots C'}_{B\ldots C} \chi^{(A}{}_E
\psi^{B\ldots C)E}_{B'\ldots C'} - m \psib^{A'B'\ldots C'}_{B\ldots C}
\chib^{E'}{}_{(B'} \psi^{AB\ldots C}_{C'\ldots)E'}\right\}
\Sigma_{AA'}  + \left\{
\psib^{A'B'\ldots C'}_{B\ldots C} D\psi^{AB\ldots C}_{B'\ldots C'}
\right\} \delta\Sigma_{AA'} \label{var}
\end{equation}
Using a formula from \cite{jf-1992-1} (which however
contains a misprint) we find $\delta\Sigma_a = 2 \chi^b{}_{[b}
\Sigma_{a]}$. If we now put all the $\sigma$-terms in $\chi^A{}_B$
equal to zero retaining only the $\tau$ terms we get
\begin{equation}
\delta\cL= \left\{
(m+1)\psib^{A'B'\!\!\ldots C'}_{B\ldots C} D\tau^{(A}{}_E
\psi^{B\ldots C)E}_{B'\ldots C'} - m \psib^{A'B'\!\!\ldots C'}_{B\ldots C}
D\taub^{E'}{}_{(B'} \psi^{AB\ldots C}_{C'\ldots)E'}  +
\psib^{E'B'\!\!\ldots C'}_{B\ldots C} D\psi^{EB\ldots C}_{B'\ldots C'}
\tau^{AA'}{}_{EE'}\right\} \Sigma_{AA'}.
\end{equation}
Integrating by parts and using the field equation several times gives
\begin{eqnarray}
\delta\cL &=&
-\psib^{A'B'\ldots C'}_{B\ldots C} \tau^{A}{}_E D\psi^{BC\ldots
E}_{B'\ldots C'}\Sigma_{AA'}
+ m\psib^{A'B'\ldots C'}_{B\ldots C}
\taub^{E'}{}_{B'}  D\psi^{AB\ldots C}_{\ldots C'E'}\Sigma_{AA'} \\
&& \qquad + \psib^{A'B'\ldots C'}_{B\ldots C}
\tau^{E}{}_{A}  D\psi^{AB\ldots C}_{B'\ldots C'}\Sigma_{EA'}
+ \psib^{A'B'\ldots C'}_{B\ldots C}
\taub^{E'}{}_{A'}  D\psi^{AB\ldots C}_{B'\ldots C'}\Sigma_{AE'}
\end{eqnarray}
Now the first and third term cancel while the second and fourth term
combine to a multiple of the field equation. Hence the functional
derivative of the action with respect to the $\tau$ terms vanishes.
By a similar argument one can show that the terms proportional to
$\sigma$ also vanish so that one has to consider only the trace free
parts proportional to $\sigma_{ABA'B'}$. In this case the calculation
is similar but more complicated, so we only state the result. The
energy momentum tensor of the spinor fields subject to equation
(\ref{feq}) is
\begin{eqnarray}
T^{ABA'B'} &=& \Im\left\{(2m-1)\psib^{C'\ldots D'(A'}_{C\ldots D}
\lambda^{B')BAC\ldots D}_{C'\ldots D'}\right.
 - \frac{m(2m+1)}{m+1} \psib^{A'B'C'\ldots D'}_{C\ldots D}
\mu^{ABC\ldots D}_{C'\ldots D'}\nonumber \\
&&\left.- \frac{m^2(2m+7)}{(m+1)(m+2)} \psib^{A'B'C'\ldots
D'(A}_{C\ldots D}
\nu^{B)C\ldots D}_{C'\ldots D'}\right\}.\label{emtensor}
\end{eqnarray}
By construction, it is divergence free and due to the conformal
invariance it is also trace free. Note, that it is made up of the fields
and all the non vanishing irreducible parts of its first derivative.
The case $m=0$ agrees with the conventional energy momentum tensor for
the Weyl equation.

\section{The Cauchy problem}

In this section we will prove that equation $(\ref{feq})$ has a well-posed
Cauchy problem, i.e., we will show that given appropriate Cauchy data
on a spatial hypersurface $S$ there will exist a unique solution of
equation $(\ref{feq})$ on a small enough neighbourhood of $S$. So existence
and uniqueness will hold (only) locally in time.

We will first examine the hyperbolicity properties of equation (\ref{feq}).
Let us write the field equation in the form
\begin{equation}
\left(\delta_{(D'}^{B'}\ldots\delta_{E'}^{C'}\delta_{A')}^{P'}
\delta_{(A}^{P}\ldots\delta_{B}^{D}\delta_{C)}^{E}\right)
\del_{PP'}\spinor{\psi}=0.\label{3.1x}
\end{equation}
We abbreviate the product of $\delta$'s by $A^a_{\mub\nu}$ thus
introducing the clumped indices $a\sim AA'$ and $\nu$, $\mub$,
indicating elements of the spin space $\spinor{S}$ and its complex
conjugate dual space. Then equation (\ref{3.1x}) has the form
\begin{equation}
A^a_{\mub\nu}\del_a\psi^\nu =0.\label{3.2x}
\end{equation}
For each covector $p_a$, $p_a A^a_{\mub\nu}$ defines a map $P$  from
$S^\nu$ into $S_\mub$ which is easily seen to define a sesquilinear
form on $S^\nu$. Before proceeding further, we will prove two useful
lemmas concerning the map $P$.
\begin{lem}\label{lem3.1}
 The map $P$ is an anti-isomorphism if and only if $p_a$
is not a null vector.
\end{lem}
\noindent {\it Proof:\/} For the sufficiency we use induction on the number
$m$ of primed indices. For $m=0$ the map is $\nu^A\mapsto
p_{A'A}\nu^A$. If $p_{A'A}\nu^A=0$ then $p^{A'}_Dp_{A'A}\nu^A=-\half
p^2\nu_D = 0$, where $p^2=p_ap^a$. Hence, if $p^2\neq 0$ then $P$ is
injective and therefore bijective. Now suppose the statement is true
for an integer $m-1$; we will show that this implies that it is also
true for $m$. In this case the map is $\spinor{\lambda}\mapsto
p_{A(A'} \lambda^{AB\ldots C}_{B'\ldots C')}$. If $p_{A(A'}
\lambda^{AB\ldots C}_{B'\ldots C')}=0$ then also $0= p^{A'}{}_B
p_{A(A'} \lambda^{AB\ldots C}_{B'\ldots C')}=\tfrac{m}{m+1}p_{A(B'}
\lambda^{AB\ldots C}_{\ldots C')A'}p^{A'}{}_B$. The induction
hypothesis implies that $p_{A}{}^{B'} \lambda^{AB\ldots C}_{B'\ldots
C'} =0$ and therefore $p_{AA'} \lambda^{AB\ldots C}_{B'\ldots C'}=0$
but this implies $p^{A'D}p_{AA'} \lambda^{AB\ldots C}_{B'\ldots C'}=
\half p^2 \lambda^{AB\ldots C}_{B'\ldots C'}=0 $. So, if $p^2\neq 0$
the map is injective and therefore bijective. If $p^2=0$ then we may
write $p_{AA'}=p_A p_{A'}$ for some spinor $p_A$ and its complex
conjugate and then $\lambda^{AB\ldots
C}_{B'\ldots C'} = p^Ap^B\ldots p^Cp_{B'}\ldots p_{C'}$ is a
nonvanishing spinor that is mapped to zero. \qed
\begin{lem}\label{lem3.2}
The determinant of the sesquilinear form defined by $P$ is
$\det(p_a A^a_{\mub\nu})=c \left(p_ap^a\right)^\tfrac{N}{2}$ where
$N=(m+1)(m+2)$ is the dimension of $S^\nu$ and $c$ is a non-zero real
number depending on the choice of basis.
\end{lem}
\noindent {\it Proof:\/} Consider the ``characteristic polynomial''
$Q(p)=\det(p_a A^a_{\mub\nu})$. As a determinant it is a Lorentz
scalar and since the only available scalar for a given $p_a$ is
$p_ap^a$ it follows that $Q(p)$ is a function of $p_ap^a$. Since
$Q(p)$ is a homogeneous polynomial in $p_a$ of degree $N$ the result
follows. Another way (which is useful later) to see this is the
following: The determinant is proportional to $\epsb^{\mub_1\ldots
\mub_N} p_a A^a_{\mub_1\nu_1}\ldots p_a A^a_{\mub_N\nu_N}
\eps^{\nu_1\ldots \nu_N}$ where $\eps^{\nu_1\ldots\nu_N}$ is a
(dual) volume form on $S^\nu$. As such it is built up from the volume
form $\eps_{AB}$ of spin space. After contracting away all the
$\eps$'s in the expression we are left with $N$ $p_a$'s which are
all contracted with each other. Due to the identity $p_{AA'}p^{AB'} =
\half p_ap^a \eps_{A'}{}^{B'}$ each pair of $p_a$'s contributes
one factor $p_ap^a$ to the determinant. Since there are $N/2$ pairs,
the result follows.\qed

{}From these two lemmas, we see that the system (\ref{3.2x}) is not
symmetric hyperbolic unless $m=0$. If this were the case, then there
would exist a timelike future pointing covector $p_a$ such that
$p_aA^a_{\mub\nu}$ was hermitian and positive definite. Although the
system is symmetric, it is not definite. For let $p_a$ be any timelike
future pointing covector with $p_ap^a=2$ and choose a spin frame
$(o^A,\iota^A)$ such that $p_a=o_Ao_{A'}+\iota_A\iota_{A'}$. Then we
choose $\spinor{\lambda}=o^Ao^B\ldots o^Co_{B'}\ldots o_{C'}$ and find
that $\bar\lambda^\mub p_a A^a_{\mub\nu} \lambda^\nu=0$ if $m>0$.
This, of course, implies that not all (real and nonvanishing)
eigenvalues of $p_aA^a_{\mub\nu}$ can be of strictly one sign. In the
Weyl case it is well known that the equation can be written in a
symmetric hyperbolic way.

To proceed further we determine the characteristics of the system
(\ref{3.2x}). These are surfaces locally described by the vanishing of
a function $\phi$ such that it is not possible to determine the
outward derivatives of a function from given Cauchy data on the
surface. Hence, on these surfaces the ``characteristic equation''
$\det(A^a_{\mub\nu}\del_a \phi)=0$ holds. Because of
lemma~(\ref{lem3.2}) each
characteristic surface is a null surface. At each point of $M$ the
normal characteristic cone defined by $Q(p)=0$ coincides with the null
cone at that point. However, unless $m=0$ the characteristic cone has
multiple sheets which implies that the system is not strictly
hyperbolic. The theory for non-strictly hyperbolic differential
operators is not as well developped as for strictly hyperbolic or
symmetric hyperbolic operators. However, in our case we can apply a
theorem of Leray and Ohya \cite{LerayOhya-1970} on non-strictly
hyperbolic systems of partial differential equations. Their main
assumption is that the characteristic determinant $Q(p)$ factorizes
such that each factor is a strictly hyperbolic polynomial%
\footnote{A polynomial $Q(\xi)$ of degree $n$ is strictly
hyperbolic iff the cone $C=\{\xi:Q(\xi)=0\}$ has a nonempty interior
such that each line through an interior point not including $\xi=0$
intersects $C$ in
exactly $n$ distinct real points.} which certainly is the case here
because $p_ap^a$ is a strictly hyperbolic polynomial of degree 2.
They show that for Cauchy data on an initial nowhere characteristic
surface $S$ which belong to a Gevrey class of functions
with index $\alpha$ the system has a unique solution in that class.
This solution admits a domain of dependence, i.e., the value of the
solution at a point depends only on the Cauchy data in the past of
that point.

Let $S$ be a spacelike hypersurface in $M$ and $t^a$ a timelike vector
field on $M$. Define a time function $t$ on $M$ by the requirement
that $t=0$ on $S$ and that $t^a\del_a t=1$. We define spacelike
surfaces $S_t$ as the surfaces of constant $t$. Given coordinates
$(x^1,x^2,x^3)$ on $S$ we can continue them off $S$ along $t^a$ by Lie
transport, i.e., by requiring that the lines $x^i={\rm const.}$ are
the integral curves of $t^a$. Thus, we obtain a coordinate system
$(t=x^0,x^1,x^2,x^3)$ on an open submanifold $\Sigma$ of $M$ that is
topologically $S\times\RR$. We choose a spin frame $(o^A,\iota^A)$
such that $o^Ao^{A'} + \iota^A\iota^{A'}= \sqrt2 n^a$, the unit normal
to $S_t$. Then $\psi_{AB\ldots CB'\ldots C'}$ has components
$\psi^\alpha =
\psi_{kk'}$ with $\alpha= (m+1) k + k'$, $0\le k < m+1$, $0\le k'
< m$ (such that $0\le\alpha < (m+1)(m+2)$) and
\begin{equation}
\psi_{kk'}=\psi_{A\ldots BC\ldots DA'\ldots B'C'\ldots D'}
\underbrace{o^A\ldots o^B}_k\iota^C\ldots \iota^D
\underbrace{o^{A'}\ldots o^{B'}}_{k'}\iota^{C'}\ldots \iota^{D'}.
\end{equation}
By taking components of equation (\ref{feq}) we obtain a system of
equations of the form ($D_a$ denoting the partial derivative with
respect to $x^a$)
\begin{equation}
G^{a\alpha}_\beta(t,x)D_{a}\psi^\beta + \Gamma^\alpha_\beta(t,x)
\psi^\beta = 0 .\label{coordeq}
\end{equation}
Here the functions $G^{a\alpha}_\beta(t,x)$ are functions on $M$ which
are algebraic expressions in the metric components $g_{ab}$ with
respect to the coordinates $(t,x)=(x^0,x^1,x^2,x^3)$. The functions
$\Gamma^\alpha_\beta(t,x)$ are algebraic expressions in the
coefficients of the spin connection. We define $a^\alpha_\beta(x,D)$ as
the linear differential operator in (\ref{coordeq}). Then we consider
the following Cauchy problem:
\begin{eqnarray}
&&a^\alpha_\beta(x,D)\psi^\beta(x)=0,\nonumber\\
&&\psi^\beta|_S \quad\hbox{is given}.
\label{Cauchy}
\end{eqnarray}

It follows from lemma~(\ref{lem3.2}) above that up to operators of lower order
than $N=(m+1)(m+2)$ we have $\det(a^\alpha_\beta(x,D))
=\dal^{\frac{N}{2}}$ where we define the determinant of noncommuting
quantities by the usual formula
$\det(a^\alpha_\beta(x,D))=\sum_{\pi\in S_N}{\rm sign}(\pi)\;
a^1_{\pi(1)}\ldots a^N_{\pi(N)}$. $\dal$ is the wave operator with
respect to the metric $g_{ab}$ expressed in the coordinates $(x^a)$.
This operator is strictly hyperbolic with respect to the hypersurfaces
$S_t$.  We are now in a position to prove the
\begin{thm} Let $\alpha$ be a real number with $1\le \alpha\le
\frac{N}{N-2}$. If the metric coefficients are in the Gevrey class
$\gamma^{\frac{3}{2}N,(\alpha)}_\infty(\Sigma)$ and if the initial data
$\psi^\beta$ are in the Gevrey class $\gamma^{(\alpha)}_2(S)$ then in
a sufficiently narrow strip $\Sigma'=\{(t,x) : 0\le t\le T\}$ around
$S$ the Cauchy problem (\ref{Cauchy}) has a unique solution
$\psi^\beta \in
\gamma^{1+\frac{N}{2},(\alpha)}_2(\Sigma)$, whose support is
contained in the domain of influence of the support of the initial
data. $\Sigma'=\Sigma$ if $1\le \alpha < \frac{N}{N-2}$.
\end{thm}
\noindent{\it Proof:\/} This is a straightforward application of
theorems of existence and uniqueness in $\S6$ of \cite{LerayOhya-1970}. We
only need to determine the various integers needed in the theorem. We
associate the integers $m^\beta=1$ with each unknown function
$\psi^\beta$ and the integers $n^\alpha=0$ with each of the equations
such that ${\rm order}(a^\alpha_\beta(x,D))\le m^\beta-n^\alpha=1$.
Then $m=\sum_\beta (m^\beta-n^\beta)=N$ is the order of $\det
(a^\alpha_\beta(x,D))$. Each of the factors in the principal part in
$\det (a^\alpha_\beta(x,D))$ is equal to the wave operator such that
$a_j=\dal$, $m_j=2$ for $(j=1,\ldots,N)$ and the number of factors is
$p=N/2$. Since in our case $r=1$ we need to add to each of the
integers $m^\beta$ and $n^\alpha$ the same integer $N/2$ in order to
satisfy the chain of inequalities
\begin{equation}
0\le r\le p\le n\le n^\alpha\le\bar{n}, n\le m^\beta, p\le m,
\end{equation}
as required in \cite{LerayOhya-1970}.  With our choice of the integers we
have $m^\beta=1+(N/2)$, $n^\alpha=0$, $p=(N/2)=n=\bar{n}$ and the
inequalities are satisfied.  According to the theorem the index
$\alpha$ of the appropriate
Gevrey classes lies in the interval $1\le \alpha \le {N\over N-2}$.
The coefficients of $a^\alpha_\beta$ are assumed
to be in $\gamma^{3N/2-1,(\alpha)}_\infty(\Sigma)$, those of
$a_j$ are in $\gamma^{j+N/2,(\alpha)}_\infty(\Sigma)$ for $0\le j\le
N/2-1$. Since all the factors are the same, $a_j=\dal$, this implies
that the coefficients are in
fact in the smallest possible class which is
$\gamma^{N-1,(\alpha)}_\infty(\Sigma)$. The coefficients of the
operator $a$ have to be in the class
$\gamma^{N/2,(\alpha)}_\infty(\Sigma)$. Taking all this together and
remembering how the coefficients in the operators are constructed from
the metric this implies that the metric coefficients have to be in
$\gamma^{3N/2,(\alpha)}_\infty(\Sigma)$. Then the conclusion of the
theorem implies that the solution of the Cauchy problem is in the
Gevrey class $\gamma^{1+N/2,(\alpha)}_\infty(\Sigma)$. \qed

According to this theorem there exists a strong correlation between
the spin of the field and the degree of smoothness of the space-time
that admits a solution of the equation. The higher the spin, the
``smoother'' the space-time has to be. The smoothness is controlled by
the number of components of the field, $N$, which depends
quadratically on the spin $m+\half$ of the field. We can improve on
this relationship somewhat by using a simplification
due to Bruhat \cite{Bruhat-1966} which is based on the observation
that if all the
minors in $\det(a^\alpha_\beta)$ have a common factor then this factor
can be ignored which results in a reduction of the number $p$ of
factors of $\det(a^\alpha_\beta(x,D))$ and therefore in the Gevrey
index $\alpha$. To be more precise, we need to prove the
\begin{lem}\label{lem3.4}
 In the adjoint matrix of $p_aA^a_{\mub\nu}$ all the entries
have the factor $(p_ap^a)^{m(m+1)/2}$ in common.
\end{lem}
\noindent {\it Proof:\/} As in the proof of lemma~(\ref{lem3.2}) this result
can be obtained by ``index counting''. The adjoint matrix is given by
the expression $\epsb^{\mub\mub_2\ldots
\mub_N} p_{a_2} A^{a_2}_{\mub_2\nu_2}\ldots p_{a_N}
A^{a_N}_{\mub_N\nu_N}
\eps^{\nu\nu_2\ldots \nu_N}$, homogeneous of degree $N-1$ in $p_a$. If
we contract over all the indices contained in all the $\eps$'s in the
volume forms we are left with an expression that contains $N-1=
(m+1)(m+2)-1$ $p_a$'s, each with one unprimed and one primed spinor
index
and has $(2m+1)$ free indices of either kind. Therefore, $(N-1)-(2m+1)$
of the $p_a$'s are contracted together, resulting in a factor
$(p_ap^a)^{m(m+1)/2}$ in each component. \qed

This lemma allows us to prove the
\begin{cor}
In the statement of the theorem we can extend the
range of the Gevrey index $\alpha$ to $1\le\alpha\le1+{1\over m}$.
More precisely, if the metric coefficients are in the Gevrey class
$\gamma_\infty^{3m+3,(\alpha)}(\Sigma)$ and if the Cauchy data are in
$\gamma_2^{(\alpha)}(S)$ then the Cauchy problem $(\ref{Cauchy})$ has
a unique solution $\psi^\beta \in \gamma_2^{m+2,(\alpha)}(\Sigma')$ in
a sufficiently narrow strip $\Sigma'$ around $S$.
\end{cor}
\noindent{\it Proof:\/} We observe that in proving the existence
theorem Leray and Ohya use a theorem for systems with diagonal
principal part (see $\S5$ of \cite{LerayOhya-1970}). To apply this theorem
one multiplies the system (\ref{feq}) with
the differential operator corresponding to the adjoint matrix of
$p_aA^a_{\mub\nu}$. This renders the principal part of the resulting
system diagonal. Due to lemma~(\ref{lem3.4}) it is enough to multiply with the
operator (of lower order) obtained from the adjoint matrix by dropping
the common factor. Then one obtains the result in a straightforward
manner by applying the theorem for diagonal systems. \qed

To end this section we want to make several remarks.
\begin{itemize}
\item[---] The case $m=0$ (the Weyl neutrino equation) is the strictly
hyperbolic case where we can choose $\alpha=\infty$. This implies that
it is possible to prescribe initial data with only a finite number of
continuous derivatives which is a known result for strictly hyperbolic
systems.
\item[---] It is also worth to mention that in all the cases there
exists a domain of influence, a fact which is taken to indicate the
hyperbolic character of partial differential equations by many authors.
\item[---] The fact that the smoothness of space-time is strongly
linked with the spin of the field is an interesting feature of this
class of equations that is not present in other spinor equations. One
has to say, though, that it is not known whether this is a
necessary consequence since the theorems only provide sufficient
conditions for existence and uniqueness.
\item[---] The diagonal system used in the proof of the corollary
corresponds to the equation $\dal^{m+1}\spinor{\psi}=0$ that
we derived in the flat case in section~2. This equation is
distinguished by the fact that it is a linear equation for
$\spinor{\psi}$ such that the coefficients are functions not of the
connection but of the curvature and its derivatives only.
\item[---] The inhomogeneous equation $\del_{A(A'}\psi^{AB\ldots
C}_{B'\ldots C')}=\chi^{B\ldots C}_{A'B'\ldots C'}$  can be treated in a
straightforward way and one obtains existence and uniqueness of
solutions in the same Gevrey class as for the homogeneous case provided
that the right hand side is in an appropriate Gevrey class, see
\cite{LerayOhya-1970}.
\end{itemize}

\section{The formal characteristic initial value problem}

In this section we want to discuss the formal aspects of the
characteristic initial value problem on a null cone for this class of spinor
equations. Due to the inherent singularity at the vertex of a null
cone this problem is very difficult to analyze and, in fact, there are
no existence results for many partial differential equations appearing
in physics, most notably the vacuum Einstein equations. So one has to
resort to formal methods to obtain at least results about the
feasability of existence theorems. A very useful method to achieve
this which is adapted to four dimensions is the method of exact sets
of spinor fields developed by Penrose \cite{Penrose-1960}. It is
based on the observation that in Taylor expansions of spinor fields
around a point it is exactly the totally symmetric derivatives of the
field that determine the restriction of the field to the null cone of
that point (the null datum). Roughly speaking, if a system of field
equations for a collection of spinor fields has the properties that
the totally symmetric derivatives are algebraically independent and if
they determine algebraically all possible derivatives of the fields
then the collection of fields is said to be exact (see
\cite{Penrose-1980}, \cite{PenroseRindlerI} for the rigorous
definition).

It has been useful to employ an algebraic formalism based on the four
derivative operators $L$, $M$, $M'$, $N$ (already mentioned in
section~2) which correspond to taking the four possible irreducible
components of the covariant derivative of an irreducible spinor field.
We will not describe the full formalism here because it would take up
too much space. Instead we will only give a brief summary and refer
for further details to \cite{jf-1993-2}. The totally
symmetric derivatives of a spinor field correspond to applying powers
of $L$ to the field. We will call an irreducible spinor to be of type
$(k,k')$ if and only if it has $k$ unprimed and $k'$ primed indices
(irrespective of their position). Acting on a spinor field of type
$(k,k')$ the operators $L$, $M$, $M'$, $N$ produce fields of
respective type $(k+1,k'+1)$, $(k+1,k'-1)$, $(k-1,k'+1)$,
$(k-1,k'-1)$. We define the operators $H$ and $H'$ by $H\phi=k\phi$
and $H'\phi=k'\phi$ for a type $(k,k')$ field $\phi$.

As we have already mentioned in section~2, the commutator of two
covariant derivatives induces commutation relations between the
derivative operators which in general involve the curvature of the
manifold and in addition the wave operator. The curvature is characterized
by three spinor fields $\Psi$, $\Phi$ and $\Lambda$ of respective
types $(4,0)$, $(2,2)$ and $(0,0)$. Before we present these relations
we need to define an algebraic operation between two spinor fields
$\phi$ and $\chi$ of respective types $(p,p')$ and $(q,q')$. The only
possible way to combine two spinor fields within the class of totally
symmetric fields in a bilinear way is by
contracting over some of the indices and then symmetrizing over the
remaining lot. This operation is entirely characterized by the numbers
of contracted primed and unprimed indices. So we define the bilinear
pairings $C_{kk'}$ by the correspondence
\begin{equation}
C_{kk'}(\phi,\chi)\cong
\phi^{A_1\ldots A_kB'_{1'}\ldots B'_{k'}}_{(C\ldots D(C'\ldots D'}
\chi_{A_1\ldots A_kB'_{1'}\ldots B'_{k'}E'\ldots F')E\ldots F)}.
\end{equation}
Then the commutation relations between the derivative operators can be
given explicitly as
\begin{eqnarray}
\left[L,N\right]&=&-(H+1)T'-(H'+1)T-\half(H+H'+2)\dal,\nonumber \\
\left[M,M'\right]&=&-(H+1)T'+(H'+1)T-\half(H-H')\dal,\label{comm} \\
\left[L,M\right]&=&-(H'+1)S, \qquad
\left[L,M'\right]=-(H+1)S',\nonumber \\
\left[N,M\right]&=& (H+1)U', \qquad \left[N,M'\right]=
(H'+1)U.\nonumber
\end{eqnarray}
The operators $S$, $T$ and $U$ and their primed versions are curvature
derivations and act on a field $\phi$ of type $(p,p')$ according to
\begin{eqnarray}
S\phi&=&p C_{10'}(\Psi,\phi) + p'C_{01'}(\Phi,\phi),\label{4.2x}\\
T\phi&=&p(p-1)C_{20'}(\Psi,\phi) + pp'C_{11'}(\Phi,\phi) -
p(p+2)C_{00'}(\Lambda,\phi),\label{4.3x}\\
U\phi&=&p(p-1)(p-2)C_{30'}(\Psi,\phi) +
p(p-1)p'C_{21'}(\Phi,\phi).\label{4.4x}
\end{eqnarray}
The action of the primed operators can be inferred from these by
formal complex conjugation. There exists an additional relation
between the derivative operators, the wave operators and the curvature:
\begin{equation}
LN - MM' = -(H'+1)T +\half H(H'+1)\dal.\label{adrel}
\end{equation}
The formulae describing the action of a derivative operator on a
bilinear pairing are quite lengthy and it is not necessary for what
follows to present them in detail (see \cite{jf-1993-2}). Symbolically
they are given by
\begin{equation}
OC(\phi,\chi)=\sum_{O} \alpha_1 C(Of,g) + \alpha_2 C(f,Og),\label{leibnitz}
\end{equation}
where $O$ is any of the derivative operators $L$, $M'$, $M$ or $N$ and
$\alpha_1$ and $\alpha_2$ are rational numbers determined by the
bilinear product and the type of $\phi$ and $\chi$.

In \cite{jf-1993-2} we showed that a collection of
fields $\{\phi_j\}$ is exact if and only if two conditions are
satisfied:
\begin{itemize}
\item[$(i)$] all the ``powers'' $L^l\phi_j$ are algebraically
independent,
\item[$(ii)$] all the ``derivatives'' of the fields, i.e., all the
expressions $s\phi_j$ where  $s$ is an arbitrary string of derivative
operators  are algebraically determined by the powers.
\end{itemize}
By algebraic independence we mean that there are no relations between
the powers involving only the bilinear pairings (and possibly the
curvature). In the same spirit we mean that the derivatives are
determined by exactly such relations in terms of the powers. Several
examples have been treated in \cite{jf-1993-3}, \cite{jf-1993-2},
\cite{PenroseRindlerI}. In a certain sense, the powers form a complete and
independent set of functions that generate the solution space of the
equations considered.

Before we consider the general case we want to study the flat case to
find the exact set structure underlying equation (\ref{feq}). So let
$\psi$ be a field of type $(m+1,m)$ satisfying the equation
$M'\psi=0$. Referring again to \cite{jf-1993-2}, we see that in that
case the expressions $L^lM^kM'{}^jN^n\dal^i\psi$ for positive integers
$l$, $k$, $j$, $n$ and $i$ generate the solution space. By use of the
field equation and the commutation relations we have $j=0$ since all
terms with $j>0$ vanish (note that in flat space $\dal$ commutes with
every derivative operator). Similarly, $i\le m$ because of
proposition~(\ref{prop2.1}) and $n+k\le m$ because $M$ and $N$ each contract
over one primed index.  Due to the additional relation (\ref{adrel})
we have the following: $L^lM^kN^n\dal^i\psi \sim L^lM^k\dal
N^n\dal^{i-1}\psi \sim L^{l+1}M^kN^{n+1}\dal^{i-1}\psi \sim \ldots
\sim L^{l+i}M^kN^{n+i}\psi$ where $a\sim b$ means ``$a$ is expressible
in terms of $b$''. So we find that in the flat case the solution space
is generated by the powers of the functions
$\psi_{ki}=M^{k-i}N^i\psi$, with $0\le k\le m$ and $0\le i \le k$.
Note that there are $\half(m+1)(m+2)$ of those functions. This number
is in agreement with the general observation that the number of null
data per point for a partial differential equation is half the number
of Cauchy data per point.

We now claim that the same set of fields is also a generating set in
the nonflat case. Before we prove this statement we need some more
preparation. Let $s_n$ denote any string of derivative operators of
length $n$. We say that $s_n$ is in normal order if and only if it has
the form $s_n=L^lM^kN^iM'{}^j$ with $l+k+i+j=n$. With each string
$s_n$ we can associate a unique normally ordered string
$\widetilde{s_n}$ of the same length in the following way: if $s_n$
does not contain $M'$ then $\widetilde{s_n}$ is the unique normally
ordered string that contains the same number of operators as $s_n$
does. If there are $M'$ operators in $s_n$ we first replace each pair
$(M,M')$ which need not be adjacent with $(L,N)$ until there is no
$M$ or $M'$ left. Then we bring the result into normal order to obtain
$\widetilde{s_n}$. Thus we get normally ordered strings containing either
$M$ or $M'$ but not both. Upon applying these normally ordered
strings to $\psi$ we get zero for all strings containing $M'$ and for
those strings with $k+i>m$. The others result in $L^lM^kN^i\psi =
L^l\psi_{k+i,i}$; these functions and their complex conjugates will be
called a normally ordered derivative or a ``power''. For each power
$L^l\psi_{ki}$ we call $l+k$ its order. Furthermore, we need to
formalize the structure of the relevant terms that will be
encountered.
\begin{defn} A {\it t-term} (``t'' for ``tree'') is recursively
defined either
\begin{itemize}
\item[$(i)$] as a power $L^l\psi_{ki}$ or $L^l\psib_{ki}$ for
nonnegative integers $l$, $k$, $i$ or
\item[$(ii)$] to be of the form $C_{kk'}(R,t)$ where $R$ is any
derivative of any of the curvature spinors and their complex
conjugates and $t$ is a t-term.
\end{itemize}
\end{defn}
\noindent A {\it t-expression} is the formal finite sum $\sum_j
\alpha_j t_j$ with coefficients $\alpha_j\in\QQ$ and t-terms $t_j$. A
t-term that is not a power will be called a {\it pure t-term}  and a
{\it pure t-expression} is a $\QQ$-linear combination of pure t-terms.

\smallbreak
The pure t-terms are binary trees with $C_{kk'}$ as nodes and with
powers or (derivatives of) curvature spinors as leaves. In fact,
exactly one leaf is a power all other ones are curvature derivatives.
This reflects the linearity of the system. We are now ready to prove
the
\begin{lem}\label{lem4.2}
 Let $\psi$ be of type $(m+1,m)$ and satisfying the
equation $M'\psi=0$. Let $s_n$ be an arbitrary string of length $n$ of
derivative operators. Then $s_n\psi$ can be written in a unique way as
$s_n\psi=\alpha\widetilde{s_n}\psi + t$ where $\alpha
\in \QQ$ and $t$ is a pure t-expression which contains only powers
of order strictly less than $n-1$.
\end{lem}
\noindent {\it Proof:\/} We use induction on the length of the string.
With $n=1$ there are four possibilities: $L\psi=L\psi_{00}$,
$M\psi=\psi_{10}$, $M'\psi=0$ and $N\psi=\psi_{11}$; so the statement
is true.

Let $O$ denote any of the derivative operators and assume the
statement to be true for all strings $s$ of length less than or equal
to $n$. Then
consider $(Os_n)\psi=O(s_n\psi)$. By the induction hypothesis and
linearity of $O$ we need to consider only two cases, namely $s_n\psi$
is $(i)$ a pure t-term or $(ii)$ a power. In case $(i)$ we need to
employ (\ref{leibnitz}) to apply $O$ to a bilinear pairing, thus
bringing $O$ inside the $C_{kk'}$ to act on each of its arguments.
Note, that then $O$ is not necessarily the same operator we started
with. When it hits the left argument, $O$ converts a curvature
derivative into a higher one thus producing a t-term of the required
type. The other argument is again either a power or a pure t-term. In
the latter case we continue descending down the tree structure until
we finally hit the power. Then we need to consider
$OL^l\psi_{ki}$. By the induction hypothesis this is a derivative of
$\psi$ of order $l+k+1 \le n-1$ and therefore equal to the sum of the
corresponding normally ordered derivative of $\psi$ and a pure
t-expression with powers of order less than $n-3$. The normally
ordered derivative is (when non vanishing) equal to a power of order
$n-1$, hence the application of $O$ to a pure t-expression yields a
pure t-expression of the required type.

In case $(ii)$ we have $s_n\psi=L^l\psi_{ki}$ with $l+k=n$. Let us
first suppose that $l\ge1$. Then
$OL^l\psi_{ki}= \left[O,L\right]L^{l-1}\psi_{ki} +
LOL^{l-1}\psi_{ki}$. In the second term we can replace
$OL^{l-1}\psi_{ki}$ with the sum of the normally ordered derivative
and a t-expression by the induction hypothesis. Then applying $L$
yields a normally ordered derivative of order $n+1$ and, as was just
shown, a t-expression of the required type. So we are left with the
commutator term. If $O=L$ we are done. If $O=M$ or $O=M'$ the
commutator term is equal to a linear combination of curvature terms by
(\ref{comm}) and (\ref{4.2x})--(\ref{4.4x}) which are t-terms of the
required type. When $O=N$ we obtain apart from curvature terms as
before a term involving the wave operator. This term can be rewritten
using (\ref{adrel}) as a linear combination of curvature terms and the
terms $LNL^{l-1}\psi_{ki}$ and $MM'L^{l-1}\psi_{ki}$. The first term
has been shown above to be of the correct type and with a similar
argument one shows that the second term is also.

Now suppose $l=0$ and $k>i$. Then we need to look at a term of the
form $OMM^{k-i-1}N^i\psi$. The only nontrivial cases are $O=M'$ and
$O=N$. In the latter case we find that the commutator term is a
curvature term and therefore of the correct type. The other term is
shown to be correct by similar arguments as above using the induction
hypothesis. In the case $O=M'$ only the wave operator term appearing
in the commutator term needs a different treatment. But this has been
shown above also to lead to correct terms.

The last case is $l=0$ and $k=i$. Then we are looking at
$ONN^{i-1}\psi$. Here all cases are trivial except for $O=N$ and this
case is treated as above. So, in summary, we have shown that all the
appearing terms are of the stated type and hence the lemma is proved.
\qed

{}From this result we can obtain a set of equations satisfied by the
functions $\psi_{ki}$. Consider $M'\psi_{ki}$, a derivative of $\psi$
of order $k+1$; therefore, there exist equations
\begin{eqnarray}
M'\psi_{ki}&=&\alpha_{ki} L\psi_{ki+1}+t'_{ki} \label{mprimepsi}\\
\noalign{\noindent and similarly}
M\psi_{ki}&=&\psi_{k+1,i}, \label{mpsi}\\
N\psi_{ki}&=&\psi_{k+1,i+1} + t_{ki}, \label{npsi}
\end{eqnarray}
where $\alpha_{ki}=0$ if $k=i$ and where $t_{ki}$ and $t'_{ki}$ are pure
t-expressions which contain $\psi_{lj}$ with $l\le k-1$ and possibly
$L\psi_{lj}$ with $l\le k-2$. If we regard these equations as the
field equations for the fields $\psi_{ki}$ then we can state the
\begin{thm} A formal solution of (\ref{feq}) gives rise to a formal
solution of (\ref{mprimepsi})--(\ref{npsi}) and vice versa. The set of
spinor fields $\{\psi_{ki}: 0\le k\le m, 0\le i\le k\}$ is exact.
\end{thm}
\noindent{\it Proof:\/} The equivalence of the two systems is obvious.
We need
to show the exactness. Here, condition $(ii)$ concerning the
completeness of the powers is an immediate consequence of the lemma.
The condition $(i)$ concerning the independence of the powers can be
verified as follows. Any relation between the powers has to be
generated by the application of the commutation relations and
(\ref{adrel}) to the field equation (\ref{feq}) and all its
derivatives. From looking at the structure of these relations one
finds that they can not link any derivatives that contain more than
two adjacent $L$'s. So all the relations that can be generated must
already be conditions on the derivatives of the field equation. But
this is a condition only on derivatives of the form $s_nM'\psi$ and
not on any power. The other possible source for conditions on powers
come from the (derivatives of the) defining equations of the
$\psi_{ki}$: $\psi_{ki}=M^{k-i}N^i\psi$. However, these are not
algebraic relations but differential relations between the functions
and --- upon taking derivatives --- between the powers. So there can
not be any relations between the powers, which therefore are
independent. \qed

This theorem shows that the characteristic initial value problem for
the equation (\ref{feq}) is formally well posed. This is, of course, a
rather weak statement, implying only that one can prescribe certain
components of derivatives of $\psi$ on the null cone of a point in
an arbitrary way and that this is just enough information for a unique
solution to exist on the level of formal power series.

The exact set $\{\psi_{ki}\}$ is not invariant (cf.
\cite{PenroseRindlerI}) because in the expressions for the derivatives
in terms of the powers there appear the curvature spinors together
with their derivatives which are taken to be known background
quantitites. Thus, these expressions depend on the actual point in
space-time that is
the vertex of the null cone. Since the field equation comes from a
variational principle and since, therefore, there exists an energy
momentum tensor we can couple the system via Einstein's equation to
the curvature. Thus, we write $G_{ab}=8\pi T_{ab}$ with $T_{ab}$ from
(\ref{emtensor}). Then we know how to express the curvature spinors
$\Phi$ and $\Lambda$ in terms of $\psi$ and its first derivatives. In
fact, $\Lambda=0$ due to the conformal invariance of the equation.
We can interpret $\psi$ as describing some kind of matter field whose
energy content creates the curvature of the manifold. We have
one more unknown function to consider, the Weyl spinor $\Psi$
which is subject to the equation (a part of the Bianchi identity)
$M'\Psi= 2M\Phi$. Referring to a theorem in \cite{jf-1993-2} we see
that the enlarged set $\{\psi_{ki},\Psi\}$ will be an invariant exact
set on $\cM$ provided that we can show that $M\Phi$ is a t-expression
and that $N\Phi=0$. We need to interpret the term ``t-expression'' a
little different now because whenever $\Lambda$ or $\Phi$ appear in
the expressions we need to substitute their resp. representations in
terms of the fields $\psi_{ki}$. Thus, we obtain an actual tree
structure built from the bilinear pairings whose leaves consist only
of powers $L^l\psi_{ki}$ and $L^l\Psi$ and their complex conjugates.
This reflects the nonlinear nature of the coupling to gravity. The
conditions above are easily verified, in fact, $N\Phi=0$ is just the
condition that the energy momentum tensor be divergence free and since
$\Phi$ itself is a t-expression its derivative is also a t-expression
as was shown above. So we have effectively proven the
\begin{thm} The set $\{\psi_{ki},\Psi\}$ subject to the equations
(\ref{mprimepsi})--(\ref{npsi}), Einstein's equation and the Bianchi
identity is an invariant exact set.
\end{thm}
\indent Thus we can make a similar statement as before concerning the
system coupled to gravity. The formal characteristic initial value
problem is well posed. In this case, we do not have a similar result
for the Cauchy problem.

\section{The general solution in Minkowski space-time}

Our aim in this section is to present the general solution of the
field equation (\ref{feq}) in flat space subject to suitable initial
and boundary conditions. Since each such solution is also a solution of
$\dal^m\psi=0$ for some positive integer $m$ we will first derive the
general solution of that equation. Since this does not depend on the
existence of spinors and on the dimension of space-time we present the
result in a slightly generalized form for arbitrary space-time
dimension. So we are working in $\MM=\RR^{1,n-1}$.  Then we will
specialize to four dimensions and restrict the kernel of $\dal^m$ to
those spinor fields that do satisfy (\ref{feq}).  Since in flat space
we are dealing with a partial differential equation with constant
coefficients the general solution could be found using methods from
the theory of distributions. We will, however, not pursue this here
but present a different approach which fits better with the
applications we have in mind.

We begin by introducing certain rings of functions associated to the
null cone of momentum space.  A function $f(k)$ defined on the null
cone in ``$k$''-space will be said to be admissible if and only if it
obeys the following conditions:
\begin{itemize}
\item[(i)] $f(k)$ is defined for all null vectors $k^a$,
\item[(ii)] $f(0) = 0$,
\item[(iii)] the function $f$ is smooth on the complement of the origin,
\item[(iv)] $\lim_{t \to 0}\, (t^r f(tk)) = 0$, for any real
number $r$ and for any non-zero null vector $k^a$; this limit must be
uniform on compact subsets of momentum space.
\end{itemize}

Condition (iv) controls both the ``infrared'' and ``ultraviolet''
behaviour of the function $f(k)$.  Condition (iv) is needed to
guarantee the differentiability and integrability of Fourier
transforms involving the function $f(k)$.

Denote by $K$, $K^+$ and $K^-$ the rings of all admissible functions,
all admissible functions that vanish identically on the past null
cone, all admissible functions that vanish on the future null cone,
respectively.  Note that we have the vector space direct sum
decomposition: $K = K^+ \oplus K^-$.

For any subspace $R$ of a ring and any vector variable $X$ denote by
$R[X]$ the space of all polynomials in the vector $X$ with
coefficients in the subspace $R$.  In particular if $R$ is itself a
(sub)-ring, then $R[X]$ is a ring.  For any non-negative integer $m$,
denote by $R_m[X]$ the subspace of the space $R[X]$ consisting of
polynomials of degree less than $m+1$ and by $R^{(m)}[X]$ the
subspace of $R_m[X]$ consisting of all polynomials homogeneous of
degree $m$ in the variable $X$.  In particular, for $x$ a space-time
vector-valued variable, we have that every element $\phi(x,k)$ of the
ring $K[x]$ has an explicit expression as a polynomial in the variable
$x$ of the following form:
\begin{equation}
\phi(x,k) = \sum_{r = 0}^{\infty} \, x^{a_1}x^{a_2}\ldots x^{a_r}
\phi_{a_1a_2\ldots a_r}(k).\label{1.1}
\end{equation}
Here each coefficient tensor, $\phi_{a_1a_2\ldots a_r}(k)$, is
completely symmetric and is an indexed element of the ring $K$.  Also
only a finite number of these coefficient tensors is non-zero.
Henceforth, each infinite sum we encounter will have only a finite
number of non-zero terms.

Denote by $\del_a$ the derivative with respect to the variable $x$ and
by $\dal \equiv g_{ab}\del^a \del^b$ the wave operator,
regarded as an endomorphism of the space $K[x]$.  Denote by $L[x]$ the
kernel of this endomorphism and define $L^+[x]$, $L_m[x]$ and
$L^+_m[x]$ as the intersections of the space $L[x]$ with the spaces
$K^+[x]$, $K_m[x]$ and $K^+_m[x]$, respectively. Consider the operator
$k_a\del^a$ as an endomorphism of $K[x]$.  Since the operators
$\dal$ and $k_a\del^a$ commute, $k_a\del^a$ restricts to an
endomorphism, denoted $D$, of $L[x]$.  Note that $D$ is the derivative
operator along the generators of the null cone restricted to solutions
of the wave equation in $K[x]$.

\subsection{The operator $D$}

\begin{prop}\label{prop5.1}
 The operator $D: L[x] \to L[x]$ is surjective provided
the space-time is at least three-dimensional.
\end{prop}
The proof of this first technical result is rather lengthy and
proceeds in several steps. We first perform a decomposition into space
and time to obtain an expression for a general element of $L[x]$.
First pick a unit timelike future pointing vector $t^a$ and denote by
$S$ the orthogonal complement in space-time of the vector $t^a$.  We
shall use lower case Latin indices from the middle of the alphabet to
label the (spatial) tensors of $S$ and shall write $\gamma_{ik}$ for
the negative of the (flat) metric induced on $S$ from the ambient
space-time metric.  Then the position vector $x^a$ decomposes as $x^a
= (t, \xi^i)$, where $t \equiv x^a t_a$ and one has the relation $x^a
x_a = t^2 - \xi^i\xi_i$, where the tensor $\gamma_{ik}$ and its
inverse are used for index lowering and raising for spatial tensors.
Correspondingly, the operators $\dal$ and $D$ decompose as $\dal =
\del_t^2 - \Delta$, where $\Delta
\equiv \gamma_{ik}\del_\xi^i \del_\xi^k$, and $D = \kappa\del_t -
(\kappa\cdot\del_\xi)$, where $k^a = (\kappa,\kappa^i)$, $\kappa \equiv
k^a t_a$ and $\kappa\cdot\del_\xi \equiv \kappa_i \del_\xi^i$.  Note
that since the vector $k^a$ is null, we have the relation $\kappa^2 =
\kappa^i\kappa_i$.


Given $\phi(x, k) \in L[x]$, define $\phi_0(\xi, k) \in K[\xi]$ and
$\phi_1(\xi, k) \in K[\xi]$ to be the restrictions to the subspace $S$
of the functions $\phi(x,k)$ and $ t_a \del^a \phi(x,k)$,
respectively.

\begin{lem}\label{lem5.2}
 The mapping $\rho_S: L[x] \to K[\xi]^2$, $\phi\mapsto
(\phi_0,\phi_1)$ is an isomorphism, mapping each solution of the wave
equation to its initial data on $S$.
\end{lem}
\noindent{\it Proof:\/} Given the pair $(\phi_0(\xi, k),\phi_1(\xi, k)) \in
K[\xi]^2$, the function $\phi(x, k)\equiv \rho_S^{-1}((\phi_0(\xi, k),
\phi_1(\xi, k))$ may be given by the following explicit formula:
\begin{equation}
\phi((t,\xi), k) = \cosh(t\Delta^{1/2})\,\phi_0(\xi,k) +
\Delta^{-1/2}\sinh(t\Delta^{1/2})\,\phi_1(\xi, k).
\label{1.2}
\end{equation}
Here the functions $\cosh(u)$ and $u^{-1}\sinh(u)$, with $u$ an
operator, are to be interpreted as formal power series.  Note that in
equation (\ref{1.2}) there are no problems with the square root of the
Laplacian, since the functions $\cosh(u)$ and $u^{-1}\sinh(u)$ are
both even.  Also there are no convergence problems, since the
functions $\phi_0$ and $\phi_1$ are polynomials in the variable
$\xi^i$.  \qed

Define the operator $\Lambda : K[\xi]^2 \to K[\xi]^2 $ by $\Lambda
\equiv \rho_S D \rho_S^{-1}$.  Then in view of lemma (\ref{lem5.2}) we
have to
show that $\Lambda$ is surjective.  We derive from equation
(\ref{1.2}) the explicit formula for the operator $\Lambda$, valid for
any pair $(\phi_0, \phi_1) \in K[\xi]^2 $:
\begin{equation}
\Lambda (\phi_0, \phi_1) = (\kappa \phi_1 -
(\kappa\cdot\del_\xi) \phi_0, \kappa \Delta\phi_0 -
(\kappa\cdot\del_\xi) \phi_1).
\label{1.3}
\end{equation}
So we must now solve the following pair of equations:
\begin{eqnarray}
 \kappa \beta - (\kappa\cdot \del_\xi) \alpha & = &\gamma,\label{1.4}\\
 \kappa \Delta\alpha - (\kappa\cdot \del_\xi) \beta & = &
 \delta.\label{1.5}
\end{eqnarray}
In equations (\ref{1.4}) and (\ref{1.5}) the pair $(\gamma, \delta)$
is a given element of the space $K[\xi]^2$ and the desired solution is
the pair $(\alpha, \beta)$ which must be shown to lie in $K[\xi]^2$.
Now it is clear from its definition that $K[\xi]^2$ is closed under
multiplication or division by $\kappa$, so we may use equation
(\ref{1.4}) to eliminate the function $\beta$ from equation
(\ref{1.5}).  This gives the following equation:
\begin{equation}
  \left(\Delta - (n\cdot\del_\xi)^2\right) \alpha = \sigma. \label{1.6}
\end{equation}
Here we have put $n\cdot\del_\xi \equiv n_i \del_\xi^i$, with $n_i
\equiv \kappa^{-1}\kappa_i$, a unit vector and $\sigma\equiv
\kappa^{-2} (\kappa\delta + (\kappa\cdot\del_\xi) \gamma) \in K[\xi]$.
Note that the desired result is false in two space-time dimensions
since the left hand side of equation (\ref{1.6}) then vanishes
identically, but the right hand side need not vanish.  So we have
reduced the problem to solving equation (\ref{1.6}), given $\sigma \in
K[\xi]$ such that the solution $\alpha$ must also lie in the space
$K[\xi]$.

\noindent{\it Proof of Proposition (5.1):\/} We first prove the
proposition for the special case with $\sigma$ of the form
\begin{equation}
\sigma = (\xi\cdot n)^p \left(\xi^2 - (\xi\cdot n)^2\right)^q
\upsilon_r. \label{1.14}
\end{equation}

Here the numbers $p$, $q$ and $r$ are non-negative integers and the
function $\upsilon_r\in K[\xi]$ is homogeneous of degree $r$ in the
vector variable $\xi^i$ and obeys both of the differential equations
$\Delta\upsilon_r = 0$ and $(n\cdot\del_\xi)\upsilon_r = 0$. It is
easy to solve equation (\ref{1.6}) in this case explicitly: a solution
is just $\alpha = ((q + 1)(n + 2r + 2q - 2))^{-1}(\xi^2 - (\xi\cdot
n)^2)\sigma$, as is easily checked, by differentiation.  This solution
clearly lies in the space $K[\xi]$ and is of the form $(\xi\cdot
n)^p \tau$, where $\tau$ statisfies the equation $(n\cdot\del_\xi)
\tau=0$.

The rest of the proof consists in a demonstration that the general
case can be reduced to this special case by decomposing $\sigma$ into
a sum of appropriate terms and then using linearity of the operator
$\Lambda$. We first decompose $\sigma$ as a sum of terms as follows:
\begin{equation}
\sigma = \sum_{r =0}^{\infty}\, {1\over r!}(\xi\cdot n)^r\sigma_r .
\label{1.7}
\end{equation}
Each coefficient $\sigma_r$ is required to obey the differential
equation $(n\cdot \del_\xi) \sigma_r = 0$.  Explicitly one has the following
formula for the quantity $\sigma_r$, valid for any non-negative
integer $r$:
\begin{equation}
\sigma_r = \sum_{s=0}^{\infty}\, {(-1)^s\over s!} (\xi\cdot n)^s
(n\cdot\del_\xi)^{r+s}\sigma. \label{1.8}
\end{equation}
In particular it is clear from equation (\ref{1.8}) that each function
$\sigma_r$ belongs to the space $K[\xi]$.  Note that the operator
$\Delta - (n\cdot\del_\xi)^2$ commutes with the multiplication
operator $(\xi\cdot n)$.  Also if $\alpha \in K[\xi]$, then we also
have $(\xi\cdot n)^r \alpha \in K[\xi]$ for any non-negative integer
$r$.  So using the linearity of equation (\ref{1.6}) and the
decomposition of equation (\ref{1.7}) and (\ref{1.8}) it suffices to
prove the solvability of equation (\ref{1.6}) with both the functions
$\sigma$ and $\alpha$ lying in the kernel of the operator
$n\cdot\del_\xi$.  Denote this kernel (a subspace of the space
$K[\xi]$) by $N[\xi]$.

Next we use a standard fact from tensor theory that any symmetric
tensor may be decomposed into tracefree parts.  In the present
language one has, for any $\tau \in N[\xi]$, a decomposition:
\begin{equation}
\tau =
\sum_{r = 0}^{\infty}\, {1\over  2^{r}r!}\left(\xi^2 - (\xi\cdot
n)^2\right)^r\tau_r. \label{1.9}
\end{equation}

Here we have put $\xi^2 \equiv \xi_i\xi^i$.  In equation (\ref{1.9}),
the coefficients $\tau_r$ must obey the differential equation:
$\left(\Delta - (n\cdot\del_\xi)^2\right) \tau_r = 0$ and must lie in
the space $N[\xi]$.  Indeed for the case that $\tau$ is a homogeneous
function of non-negative integral degree $m$ in the vector variable
$\xi$, each function $\tau_r$ may be given explicitly by the
following formula:
\begin{equation}
\tau_r =
\sum_{s = 0}^{\infty}\, (-1)^s {(\lambda - 2r)\Gamma(\lambda - 2r - s) \over
2^{r+2s} \Gamma(\lambda - r + 1) \Gamma(s+1)} \left(\xi^2 - (\xi\cdot
n)^2\right)^s (\Delta-(n\cdot\del_\xi)^2)^{r+s}\tau,  \label{1.10}
\end{equation}
with $\lambda \equiv (n - 4 + 2m)/2$. It is easily checked by
differentiation that the function $\tau_r$ of equation (\ref{1.10})
lies in the kernel of both the operators $\Delta - (n\cdot\del_\xi)^2$
and $n\cdot\del_\xi$ as required.  The proof of compatibility of
equations (\ref{1.9}) and (\ref{1.10}) follows immediately from the
lemma (\ref{lem5.3}) given below.

Since every $\tau \in K[\xi]$ is uniquely a sum of its homogeneous
parts and each of its homogeneous parts lies also in the space
$K[\xi]$, equations (\ref{1.9}) and (\ref{1.10}) hold also for
inhomogeneous functions $\tau \in N[\xi]$, provided that equation
(\ref{1.10}) is rewritten as follows:
\begin{equation}
\tau_r = \sum_{s = 0}^{\infty}\,{(-1)^s \over
2^{r+2s}} {\mu\Gamma(\mu - s)\over \Gamma(\mu + r + 1)\Gamma(s+1)}
\left(\xi^2 - (\xi\cdot n)^2\right)^s
(\Delta-(n\cdot\del_\xi)^2)^{r+s}\tau,
\label{1.13}
\end{equation}
where $\mu \equiv (n - 4 + 2\xi_A \del_\xi^A)/2$.

Combining these results and again using the linearity of equation
(\ref{1.6}), we see that it is sufficient to prove the solvability of
equation (\ref{1.6}) with the function $\sigma$ being of the form
given in (\ref{1.14}) above. This completes the proof of
proposition~(\ref{5.1}).\qed

Note that since the above proof is compatible with homogeneity, it
also shows that the restriction of the map $D$ to the subspace
$L_m[x]$ has range the subspace $L_{m-1}[x]$, for every positive
integer $m$ and that the restriction of $D$ to the subspace of
$L^{(m)}[x]$ is surjective onto the subspace $L^{(m-1)}[x]$.

\begin{lem}\label{lem5.3}
  For $j$ a non-negative integer, define a function
$g_j(z)$ of the complex variable $z$ as follows:
\begin{equation}
 g_j(z) = \sum_{r = 0}^{j}\, (- 1)^r {(z - 2r)\Gamma(z - r - j)
\over \Gamma(z - r + 1) \Gamma(r+1) \Gamma(j - r + 1)}.
\label{1.11}
\end{equation}
Then the function $g_j$ vanishes identically unless $j = 0$ and the
function $g_0$ is the constant function with value one.
\end{lem}
\noindent{\it Proof:\/} First, the case $j = 0$ is easily checked by
inspection.  So henceforth assume, for convenience, that $j$ is a
fixed positive integer. From its definition it is clear that the
function $g_j(z)$ is a rational function of the variable $z$ and that
$lim_{z \to \infty}\, g_j(z) = 0$.  Therefore, the result will follow
if it is proved that the function $g_j(z)$ is periodic.  But one has
the following relations, using equation (\ref{1.11}):
\begin{eqnarray}
g_j(z) &=& \sum_{r = 0}^{j}\, (- 1)^r {(z - r)\Gamma(z - r - j) \over
\Gamma(z - r + 1) \Gamma(r+1) \Gamma(j - r + 1)}
 - \sum_{r = 1}^{j}\,(- 1)^r {\Gamma(z - r - j) \over
\Gamma(z - r + 1) \Gamma(r) \Gamma(j - r + 1)}  \nonumber \\
& = &\sum_{r = 0}^{j}\,
(- 1)^r {\Gamma(z - r - j)\over
\Gamma(z - r) \Gamma(r+1) \Gamma(j - r + 1)}
 + \sum_{r =0}^{j-1}\,
(- 1)^r {\Gamma(z - r - j - 1) \over
\Gamma(z - r) \Gamma(r+1) \Gamma(j - r)}  \label{1.12}
\\
& = & \sum_{r = 0}^{j}\,
(- 1)^r {(z - r - j - 1)\Gamma(z - r - j - 1) \over
\Gamma(z - r) \Gamma(r+1) \Gamma(j - r + 1)} + \sum_{r = 0}^{j}\,
(- 1)^r {(j - r)\Gamma(z - r - j - 1) \over
\Gamma(z - r) \Gamma(r+1) \Gamma(j - r + 1)}  \nonumber\\
& = &\sum_{r = 0}^{j}\,
(- 1)^r {(z - 2r - 1)\Gamma(z - r - j - 1) \over
\Gamma(z - r) \Gamma(r+1) \Gamma(j - r + 1)} = g_j(z - 1). \nonumber
\end{eqnarray}
So the function $g_j(z)$, for $j > 0$ is periodic and therefore
vanishes identically as required. \qed

\subsection{The Fourier transform operator $\cF$.}
Consider the $(n-1)$-form $\Xi$ defined on the null cone in momentum
space defined by the formula $\Xi\equiv dk^{a_1}dk^{a_2}\ldots
dk^{a_{n-1}} = k_{a_0} \epsilon^{a_0a_1a_2\ldots
a_{n-1}}\Xi$. Restricted to the null cone one has $k_ak^a = k_a dk^a =
0$, so the left hand side of this equation is orthogonal to the null
vector $k^a$.  The $(n-1)$-form $\Xi$ factorizes according to the
formula $k^a\Xi = dk^a\omega$, where the $n-2$-form $\omega$ is
defined by the formula: $k^{[a_1}dk^{a_2}\ldots dk^{a_{n-1}]} =
k_{a_0} \epsilon^{a_0a_1a_2
\ldots a_{n-1}}\omega$.  Both the forms $\Xi$ and $\omega$ are closed:
$d\Xi = d\omega = 0$.

Given any $\lambda(x,k) \in K[x]$, we consider its generalized Fourier
transform $\cF(\lambda)$, which is a space-time field given by the
following formula, valid at any point $x$ of $\MM$:
\begin{equation}
 \cF(\lambda)(x) \equiv \int e^{ik_bx^b}\lambda(x,k)\,\Xi.\label{2.1}
\end{equation}
The integration in equation (\ref{2.1}) is to be carried out over the
complete (past and future) null cone, equipped with the induced
orientation from its embedding in $k$-space, which in turn is oriented
by the volume form, $\epsilon_{a_1a_2\ldots a_n} dk^{a_1} dk^{a_2}
\ldots dk^{a_n}$. Note that by definition of the space $K[x]$ the
convergence of the integral of equation (\ref{2.1}) is automatic and
the resulting field $\cF(\lambda)$ is everywhere smooth on space-time.

The linear operator $\cF: \lambda \mapsto \cF(\lambda)$ is defined on
$K[x]$ and we denote its range by $\Gamma[x]$. Also denote by
$\Gamma$, $\Gamma^+$ and $\Gamma^+[x]$ the images under the operator
$\cF$ of the spaces $K$, $K^+$ and $K^+[x]$, respectively.

It is clear that the space $\Gamma[x]$ consists of certain polynomials
in the variable $x^a$ with coefficients in the space $\Gamma$, so to
understand the range of the operator $\cF$ it is sufficient to
identify the space $\Gamma$.

To this end, we first introduce for any $\alpha$ and $\beta$,
solutions of the wave equation in spacetime the $n-1$-form
$\omega(\alpha, \beta) \equiv \alpha (*d\beta) - \beta (*d\alpha)$,
where $*$ is the Hodge star operator on forms for the given Lorentzian
metric.  Since the wave equation for a field $\phi$ may be written
$d*(d\phi) = 0$, it is clear that the form $\omega(\alpha,\beta)$ is
closed.  Define $\Omega(\alpha, \beta)$ to be the integral of the form
$\omega(\alpha, \beta)$, over a spacelike hypersurface, oriented
towards the future, asymptotic to spacelike infinity, for given fields
$\alpha$ and $\beta$, which are required to be such that the integral
converges and is independent of the choice of that hypersurface.
Denote by $W$ the space of all solutions of the scalar wave equation
with initial data, on any spacelike hypersurface, asymptotic to
spacelike infinity, in the Schwarz class (the initial data for a
solution $\phi$ on a hypersurface is by definition the restriction of
$\phi$ and $*d\phi$ to that hypersurface).  Denote by $M[x]$ the space
of all polynomial solutions of the wave equation and by $M'[x]$ its
dual space.  Then for each $\phi(x) \in W$, we obtain an element
$\mu(\phi)$ of the space $M'[x]$, defined by the formula $\mu(\phi)(f)
= \Omega(\phi, f)$, for each $f \in M[x]$.  This gives a moment map
$\mu: W \to M'[x]$, $\phi \mapsto \mu(\phi)$.  Then we have the
following result:

\begin{prop}\label{prop5.4}
$\Gamma = Ker(\mu)$.
\end{prop}
\noindent
The proof of this result follows immediately from the Fourier
inversion formula. \qed

Note that the information in the moment map $\mu$ is completely
contained in the formal power series defined by the formula:
$\rho(\phi)(p_a) = \Omega(e^{ip_ax^a}, \phi)$, where the exponential
$e^{ip_ax^a}$ is understood as a formal power series in the null
covector $p_a$.  The quantity $\rho(\phi)(p_a)$ is then a formal power
series whose coefficients are tracefree symmetric tensors,
representing the various moments of the field $\phi$.  In terms of
initial data, the quantity $\rho(\phi)$ represents all moments of the
data for the field $\phi$.  In this language the space $\Gamma$ is the
subspace of the space $W$ consisting of all fields $\phi$, for which
$\rho(\phi) = 0$.

Our final aim in this subsection is to determine the kernel of $\cF$ and to
prove the

\begin{prop}\label{prop5.5}
For $\lambda \in K[x]$ we have
\begin{eqnarray}
 \cF(\lambda)(x)=& \int e^{ik_bx^b}\lambda(x,k)\,\Xi = 0 \quad
\hbox{for all $x$} & \label{2.2}\\
&\iff \lambda(i\del_k + t k) = 0. & \label{2.3}
\end{eqnarray}
\end{prop}
Here the scalar $t$ is an indeterminate. Also in
writing equation (\ref{2.3}) it is to be understood that in each term
of the expression of the function $\lambda(x, k)$ as a polynomial in
the variable $x$, the expression is ordered by placing all the factors
of the variable $x$ to the left, before replacing the variable $x$ by
the operator $i\del_k + tk$.

Before we proceed to the proof of the proposition we want to
clarify the structure of equation (\ref{2.3}) with an example. In
the case $\lambda \in K_2[x]$, we have the expression $\lambda(x,k) =
x^ax^b\beta_{ab} + x^a\beta_a + \beta$, for some symmetric tensor
$\beta_{ab}$, vector $\beta_a$ and scalar $\beta$, each of which
depends only on the variable $k$.  For this case equation (\ref{2.3})
reads as follows:
\begin{equation}
 0= (i\del_k^a + tk^a)(i\del_k^b + tk^b)\beta_{ab} +
(i\del_k^a + tk^a)\beta_a + \beta.\label{2.4}
\end{equation}

Note that the commutator $(i\del_k^a + tk^a)(i\del_k^b + tk^b) - (i\del_k^b
+ tk^b)(i\del_k^a + tk^a)$ vanishes identically, so there is no factor
ordering problem.  Expanding equation (\ref{2.4}) in powers of the
indeterminate $t$, equation (\ref{2.4}) is equivalent to the following
three equations:
\begin{eqnarray}
 0&=& k^ak^b\beta_{ab},\label{2.5}\\
 0&=& (\del_k^ak^b + k^a\del_k^b)\beta_{ab} - ik^a\beta_a,\label{2.6}\\
 0&=& \del_k^a\del_k^b\beta_{ab} -
i\del_k^a\beta_a - \beta.\label{2.7}
\end{eqnarray}
Note that equations (\ref{2.3}) -- (\ref{2.7}) implicitly require that
one extend the function $\lambda(x,k)$ off the null cone of momentum
space before writing these equations since the formulation of the
equations uses the full derivative operator $\del_k$. However it must be
possible to rewrite the equations so that they are purely intrinsic to
the null cone. In the case of equations (\ref{2.5}) -- (\ref{2.7}),
one can see easily that these equations are equivalent to the
following three equations:
\begin{eqnarray}
 0&=& k^ak^b\beta_{ab},\label{2.8}\\
 0&=& (L^ak^b + k^aL^b)\beta_{ab} - 3p^ak^b\beta_{ab} -
     ip^ck_ck^a\beta_a,\label{2.9}\\
 0& =& L^aL^b\beta_{ab} - 3p^aL^b\beta_{ab} - ip^ak_aL^b\beta_b +
   2p^ap^b\beta_{ab} \nonumber \\
  & + & ip^ak_ap^b\beta_b + (1/2)p^ap_a(ik_b\beta^b +
   \beta_b^b) - (p^ak_a)^2\beta.\label{2.10}%
\end{eqnarray}
Here the operator $L^a$ is defined for any fixed vector $p^a$ by the
relation:
\begin{equation}
 L^a \equiv 2p_bk^{[b}\del_k^{a]}. \label{2.11}
\end{equation}
It is clear from its definition that the operator $L^a$ is intrinsic
to the null cone.  Then equations (\ref{2.8})--(\ref{2.10}) hold for
arbitrary vectors $p^a$.  Alternatively equations (\ref{2.9}) and
(\ref{2.10}) can be rewritten without using the vector $p^a$ as
follows:
\begin{eqnarray}
 0 & = &(L_c\,^ak^b + k^aL_c\,^b)\beta_{ab} - 3k^b\beta_{cb} -
ik_ck^a\beta_a,\label{2.12}\\
 0 & = &L_{(c}\,^aL_{d)}\,^b\beta_{ab} - 3L_{(c}\,^b\beta_{d)b} -
ik_{(c}L_{d)}\,^b\beta_b + 2\beta_{cd} + ik_{(c}\beta_{d)} +
(1/2)g_{cd}(ik_b\beta^b + \beta_b{}^b) - k_ck_d\beta.\label{2.13}
\end{eqnarray}
Here we have introduced the intrinsic operator $L^{ab}$, given by the
formula: $L^{ab} \equiv 2k^{[a}\del_k^{b]}$, in terms of which one has
the relation $L^a = p_bL^{ba}$.  Note that provided that equation
(\ref{2.8}) also holds, each of the equations (\ref{2.12}) and
(\ref{2.13}) amounts to just one scalar equation, since one may verify
that the right hand side of equations (\ref{2.12}) and (\ref{2.13})
are proportional to the quantities $k_c$ and $k_ck_d$, respectively.

\noindent{\it Proof of the proposition (\ref{5.5}):\/} We shall prove the
statement for $\lambda \in K_m[x]$, for every non-negative integer $m$
by induction on the natural number $m$. First the required result
holds for $m=0$, since in this case the function $\lambda(x, k) =
\beta(k)$ for some function $\beta(k)\in K$.  When equation
(\ref{2.2}) holds, the integral $ \int e^{ik_ax^a}\beta(k)\,\Xi\,$
gives the zero solution of the wave equation $\dal\Phi = 0$ and it is
well known in this case that this entails that the function $\beta$
must vanish identically.

Next suppose the required result is true for all $\lambda \in K_m[x]$
for all $m<s$, for some positive integer $s$. We prove the result for
$m=s$.  So consider equation (\ref{2.2}) with the function
$\lambda(x,k) \in K_s[x]$ now a polynomial in the variable $x$ of
degree not more than $s$.  Then we can decompose the function
$\lambda(x,k)$ as $\lambda(x,k) = \alpha(x, k) + \beta(x, k)$, where
$\alpha(x, k) \in K^{(s)}[x]$ and $\beta(x, k) \in K_{s-1}[x]$.

Applying the wave operator to equation (\ref{2.2}), we get
the following equation:
\begin{equation}
 0=\int e^{ik_bx^b}(2ik_a\del^a + \dal) \lambda(x,k)\,\Xi.\label{2.14}
\end{equation}
Since the function $(2ik_a\del^a + \dal)\lambda(x,k) \in
K_{s-1}[x]$, we get by the inductive hypothesis, the equation:
\begin{equation}
0 = \left[\left(2ik_a\del^a + \dal\right)\lambda(x, k)\right]_{x
\to(i\del_k + tk)}.\label{2.15}
\end{equation}
Next take the partial derivative of equation (\ref{2.2}) with respect to the
variable $x$.  We get the following equation:
\begin{equation}
 0=\int e^{ik_bx^b}(ik^a + \del^a)(\alpha + \beta)\,\Xi.\label{2.16}
\end{equation}
Now we have $\alpha(x,k) = x^e\alpha_e(x, k)$, where the function
$\alpha^e(x, k) \equiv s^{-1}\del^e\alpha(x, k) \in K_{s-1}[x]$.
Then equation (\ref{2.16}) may be rewritten, using an integration by
parts as follows:
\begin{eqnarray}
0 & = &\int e^{ik_bx^b}(ik^ax^e\alpha_e + ik^a\beta +
\del^a\lambda)\,\Xi \nonumber \\
& = &\int e^{ik_bx^b}(2ik^{[a}x^{e]}\alpha_e + ik^a\beta +
\del^a\lambda)\,\Xi \,+\, \int
e^{ik_bx^b}ix^ak^e\alpha_e\,\Xi \nonumber \\
& = &\int e^{ik_bx^b}(-2k^{[a}\del_k^{e]}\alpha_e + ik^a\beta +
\del^a\lambda)\,\Xi \,+ \, x^a\int
e^{ik_bx^b}ik^e\alpha_e\,\Xi.\label{2.17}
\end{eqnarray}
Now using equation (\ref{2.14}), we have the following:
\begin{equation}
0=\int e^{ik_bx^b}(2ik_a\del^a + \dal)\lambda\, \,\Xi
 = \int e^{ik_bx^b}(2isk^a\alpha_a + 2ik_a\del^a\beta +
\dal\lambda)\,\Xi. \label{2.18}
\end{equation}
Hence one has the following equation:
\begin{equation}
\int e^{ik_bx^b}ik^a\alpha_a\,\Xi = -{1\over 2s} \int e^{ik_bx^b}
(2ik_a\del^a\beta + \dal\lambda)\,\Xi. \label{2.19}
\end{equation}
This equation is used to replace the last integral of equation
(\ref{2.17}).  We then obtain
\begin{equation}
0 =\int e^{ik_bx^b}(-2k^{[a}\del_k^{e]}\alpha_e + ik^a\beta +
\del^a\lambda) - {x^a\over 2s}
(2ik_a\del^a\beta + \dal\lambda)\,\Xi.\label{2.20}
\end{equation}
Now, by inspection, each term multiplying the quantity $e^{ik_bx^b}$
of equation (\ref{2.20}) is of degree at most $s-1$ in the variable
$x$.  Therefore we may invoke the inductive hypothesis again to deduce
\begin{eqnarray}
 0 &=&\left[-2k^{[a}\del_k^{b]}\alpha_b + ik^a\beta +
   \del^a\lambda - {x^a\over 2s}(2ik_a\del^a\beta
   + \dal\lambda)\right]_{x\mapsto (i\del_k + tk)}\nonumber \\
 &=&\left[2ik^{[a}x^{b]}\alpha_b + ik^a\beta + \del^a\lambda -
   {x^a\over 2s}(2ik_a\del^a\beta +  \dal\lambda)\right]_{x
\mapsto (i\del_k + tk)}. \label{2.21}
\end{eqnarray}
Now equation (\ref{2.14}), when written out gives the equation:
\begin{equation}
\left[2ik_a\del^a\beta + \dal\lambda\right]_{x
\mapsto (i\del_k + tk)} = \left[- 2ik_a\del^a\alpha\right]_{x
\mapsto (i\del_k + tk)} = \left[- 2isk_a\alpha^a\right]_{x \mapsto
(i\del_k + tk)}.
\label{2.22}
\end{equation}
Substituting equation (\ref{2.22}) into the last part of equation
(\ref{2.21}) gives the following equation:
\begin{eqnarray}
0 & = &\left[2ik^{[a}x^{b]}\alpha_b + ik^a\beta +
\del^a\lambda + ix^ak_b\alpha^b\right]_{x \mapsto (i\del_k +
tk)}\nonumber \\
& = &\left[ik^ax^b\alpha_b + ik^a\beta +
\del^a\lambda\right]_{x \mapsto
(i\del_k + tk)} =\left[ik^a\alpha + ik^a\beta +
\del^a\lambda\right]_{x \mapsto (i\del_k + tk)}\label{2.23}\\
& = &\left[i(k^a - i\del^a)\lambda\right]_{x \mapsto (i\del_k + tk)}
 = ik^a \left[\lambda\right]_{x \mapsto (i\del_k + tk)}.\nonumber
\end{eqnarray}
In the transition from the penultimate to the last line of equation
(\ref{2.23}), we have used the fact that the terms arising from the
commutator of the operator of multiplication by $k^a$ and the operator
$i\del_k + tk$ exactly cancel the derivative term, the quantity
$i\del^a\lambda$.  That this is correct may be seen as follows.
Consider the quantity $\left[(k^a - i\del^a) \left( (x^bp_b)^n f(k)
\right) \right]_{x \mapsto (i\del_k + tk)}$, for $n$ a non-negative
integer, $p$ a constant covector and for $f\in K$.  We then have the
following equation:
\begin{eqnarray}
&& \left[(k^a - i\del^a)\left((x^bp_b)^nf(k)\right)\right]_{x
\mapsto (i\del_k + tk)} = \left[(x^bp_b)^n k^af(k) -
inp^a(x^bp_b)^{n-1}f(k)\right]_{x \mapsto (i\del_k + tk)} \nonumber\\
& =& ((i\del_k^b + tk^b)p_b)^n k^af(k) -
\left[inp^a(x^bp_b)^{n-1}f(k)\right]_{x \mapsto (i\del_k + tk)} \nonumber\\
& = &k^a\left((i\del_k^b + tk^b)p_b\right)^n f(k) + inp^a
\left((i\del_k^b + tk^b)p_b\right)^{n-1} f(k) -
inp^a\left[(x^bp_b)^{n-1}f(k)\right]_{x \mapsto (i\del_k + tk)} \nonumber\\
& = &k^a \left[(x^bp_b)^n f(k)\right]_{x \mapsto (i\del_k + tk)}.
\label{2.24}
\end{eqnarray}
Since any polynomial $\lambda \in K[x]$ may be written as a finite
linear combination of terms of the form $(x^bp_b)^nf(k)$, we have the
relation $[(k^a - i\del^a)\lambda]_{x \mapsto (i\del_k + tk)} =
k^a [\lambda]_{x \mapsto (i\del_k + tk)}$, for any $\lambda \in K[x]$,
as required.

Finally we remove the factor $ik^a$ from equation (\ref{2.23}) giving
the required result and the induction is complete. \qed

\subsection{The general solution of $\dal^m\psi=0$}

In this subsection we provide the general solution of the equation
$\dal^m \psi = 0$ in the space $\Gamma[x]$.  As the construction will
show this is equivalent to finding the general solution subject only
to the condition that the zrm field $\dal^{m-1} \psi$ lies in the
space $\Gamma$ (i.e. has zero moment map).  We begin by proving that
there is no loss of generality in restricting the domain of the
Fourier transform operator $\cF$ from the space $K[x]$ to its subspace
$L[x]$.  More specifically, we have

\begin{prop}\label{prop5.6}
For each $\lambda(x,k) \in K[x]$ there exists a
$\mu(x,k) \in L[x]$ such that $\cF(\lambda)=\cF(\mu)$.
\end{prop}

\noindent{\it Proof:\/} For any $\lambda \in K[x]$, we have
the decomposition, directly analogous to that of equation (\ref{1.13})
above:
\begin{eqnarray}
\lambda & = & \sum_{r = 0}^{\infty}\,{(x^2)^r\over 2^r r!}\lambda_r.
        \label{3.1}\\
\lambda_r & = & \sum_{s = 0}^{\infty}\,{(-1)^s{x^2}^s\over 2^{2s+r}}
{\Gamma(\nu - 2r - s) (\nu - 2r) \over \Gamma(\nu - r + 1) \Gamma(s+1)}
\dal^{r+s}\lambda.\label{3.2}
\end{eqnarray}
Here we have put $x^2 \equiv x_ax^a$ and $\nu \equiv (n - 2 +
2x^a\del_a)/2$.  From equation (\ref{3.2}), by differentiation, it
follows that the coefficients $\lambda_r$ belong to the space $L[x]$.
In particular it follows that the space $K[x]$ is the sum of the space
$L[x]$ with the module generated over the ring $K[x]$ by the function
$x^2$.  In view of this decomposition, to prove the required result it
suffices to show that for any function $\alpha \in K_m[x]$ there
exists a function $\beta \in K_{m+1}[x]$ such that $\cF(x^2\alpha) =
\cF(\beta)$.  Now we have the integration by parts identity, valid for
any $y_a \in K[x]$, such that $y_ak^a = 0$:
\begin{equation}
0 = \int \del^a(e^{ik_bx^b}y_a)\,\Xi. \label{3.3}
\end{equation}
Rephrasing equation (\ref{3.3}) in terms of the operator $\cF$, we
have $\cF(\del^ay_a + ix^ay_a) = 0$, valid for any $y_a \in K[x]$,
such that $y_ak^a = 0$.  In particular consider the case that $y_a
\equiv - ik_0^{-1}(k_0 x_a - t_a x^b k_b + k_a x^bt_b)\alpha$, where
$t^a$ is a fixed unit vector and $k_0 \equiv k^a t_a$.  Then it is
clear that $y_a \in K_{m+1}[x]$ and that $y_a$ satisfies the identity
$y_ak^a = 0$, so one has $\cF(- \del^ay_a) = \cF(ix^ay_a)$.  But by
contracting the vectors $x^a$ and $iy_a$ we also have the relation
$ix^ay_a = x^2\alpha$, which yields the relation $\cF(x^2\alpha) =
\cF(\beta)$, where $\beta \equiv - \del^ay_a$.  Since from its
definition it is clear that the function $\beta$ lies in the space
$K_{m+1}[x]$, the proof is complete.  \qed

In view of this result we henceforth assume without loss of generality
that $\cF$ is defined on the space $L[x]$.

The wave operator of space-time, $\dal$ acts naturally on the space
$\Gamma[x]$ and one has the relation, immediate from the definition of
$\cF$ in equation (\ref{2.1}), valid for any $\lambda \in L[x]$:
\begin{equation}
\dal \cF(\lambda) = 2i\cF(D\lambda).
\label{3.4}
\end{equation}
We next wish to determine the kernel of the operator $\dal$ acting on
the space $\Gamma[x]$.  By equation (\ref{2.2}) above and using
equation (\ref{3.4}), we have $\dal\cF(\lambda) = 0$, for $\lambda \in
L[x]$ if and only if the function $\lambda(x, k)$ obeys the equation:
\begin{equation}
 0 = (D \lambda)(i\del + tk, k). \label{3.5}
\end{equation}
Let the power series expansion of the function $\lambda \in L[x]$ be
given as follows:
\begin{equation}
\lambda(x,k) = \sum_{r = 0}^{\infty}\, {1 \over r!} x^{a_1} x^{a_2}
\ldots x^{a_r}\lambda_{a_1a_2\ldots a_r}(k). \label{3.6}
\end{equation}
In equation (\ref{3.6}), each coefficient tensor
$\lambda_{a_1a_2\ldots a_r}$ belongs to the space $K$ and is
completely symmetric and tracefree.  Writing out equation (\ref{3.5})
in terms of this expansion gives the following system of equations,
one for each positive integer $q$:
\begin{equation}
0 = \sum_{p = 0}^{\infty}\,{1 \over p!} \lambda_{p,q}(k). \label{3.7}
\end{equation}
Here the quantity $\lambda_{p, q}$ is by definition $\lambda_{p,
q}\equiv i^p\del^{a_1}\ldots \del^{a_p}k^{b_1}\ldots
k^{b_q}\lambda_{a_1\ldots a_pb_1\ldots b_q}$.  Note that there are no
factor ordering problems for the quantity $\lambda_{p,q}$, since the
tensor coefficients are all tracefree.  Now expanding in powers of the
indeterminate $t$ we have the identity, derived from equation
(\ref{3.6}):
\begin{equation}
\lambda(i\del + tk, k) = \sum_{p, q = 0}^{\infty}\,
{t^q \over p! q!} \lambda_{p,q}(k). \label{3.8}
\end{equation}
Comparing equations (\ref{3.8}) and (\ref{3.7}), we see that if we
define $\mu(x, k) \equiv \lambda(x, k) - \lambda(i\del, k)$, then we
have the relation:
\begin{equation}
0 = \mu(i\del + tk, k). \label{3.9}
\end{equation}
Note that $\mu \in L[x]$, so from equation (\ref{3.9}), one has $\mu
\in \ker(\cF)$.  This gives the relation $\cF(\lambda) = \cF(\nu)$,
where $\nu\equiv \lambda(i\del, k)\in L_0[x]$.  So we have shown that
if $\dal \cF(\lambda) = 0$, then $\lambda = 0 \bmod (\ker(\cF) +
L_0[x])$.  Conversely, if $\lambda \in \ker(\cF) + L_0[x]$, then it is
clear from equation (\ref{3.4}) that $\dal \cF(\lambda) = 0$.  So we
have proved the relation $\ker(\dal \cF) = \ker(\cF) + L_0[x]$.
Rephrasing we have proved the relation $ \ker(\dal) = \cF(L_0[x])$.

This generalizes immediately to our first main result
\begin{thm}\label{thm5.7}
 For any integer $m > 0$ and any space-time dimension $n>2$, the
kernel of the operator $\dal^m$ when acting on $\Gamma[x]$ is
given by the relation
\begin{equation}
\ker(\dal^m \cF) = \ker(\cF) + L_{m - 1}[x].\label{3.10}
\end{equation}
Equivalently this may be stated as:
\begin{equation}
\ker(\dal^m) = \cF(L_{m-1}[x]).\label{3.11}
\end{equation}
\end{thm}
\noindent{\it Proof:\/} The required result has just been proved in the case $m
= 1$, so henceforth assume $m$ is a fixed integer greater than one.
Suppose that $\Phi\in \ker(\dal^m)$ and $\Phi = \cF(\phi)$, for some
$\phi \in L[x]$.  Then $\dal^{m-1}\Phi \in \ker(\dal)$, so we have
$\dal^{m-1}\Phi = \cF(\alpha)$, for some $\alpha \in L_0[x]$.  By
proposition (\ref{5.1}) the operator $D$ is surjective as a
linear map from the space $L_{s}[x]$ to the space $L_{s - 1}[x]$, for
any positive integer $s$.  It immediately follows that the operator
$D^r$ is also surjective as a linear map from the space
$L_{s}[x]$ to the space $L_{s - r}[x]$, for any positive integers $r$
and $s$, with $s$ not less than $r$. Therefore we may put $\alpha(x,
k) = (2i D)^{m - 1} \beta(x, k)$ for some $\beta \in
L_{m-1}[x]$.  Then we have $\dal^{m-1}\Phi = \cF(\alpha) =
\cF\left((2i D)^{m - 1} \beta\right) =
\dal^{m-1}(\cF(\beta))$.  So we have $\Phi - \cF(\beta) \in
\ker(\dal^{m - 1})$ and the required result now follows immediately by
induction. \qed

We have shown that the general solution of the equation
$\dal^{m+1}\phi=0$ in the space $\Gamma[x]$ is given by an integral
formula
\begin{equation}
\phi(x) = \int e^{ik_bx^b} \left\{\lambda(k) + \sum_{i=1}^m
\lambda_{a_1\ldots a_i}(k) x^{a_1} \ldots x^{a_i}\right\} \,\Xi
\end{equation}
where the polynomial inside the integral satisfies the wave equation.
The solutions of this type are automatically $\cC^\infty$. So with
this formula we can not reach functions which are merely $\cC^k$
differentiable. However, this class of solutions is sufficient for our
purposes. More general solutions can be obtained using functional
analytic methods by starting with this integral formula on an
appropriate function space and then taking limits. We will not pursue
this here.

\subsection{Spinor momentum space}

We now specialize to the case of four dimensions and introduce spinor
variables.  A future pointing null momentum covector $k_a$ may be
factorized as $k_a = k_A k_{A'}$, for $k_A$ a two component spinor,
with complex conjugate spinor $k_{A'}$.  More precisely, we have a
surjective map from the momentum spin space to the future null cone of
momentum space, which maps the spinor $k_A$ to the null vector $k_A
k_{A'}$.  The inverse image of the vector $k_a \equiv k_A k_{A'}$ is
the circle of spinors $\alpha k_A$, (with $\alpha \in \CC$ and
$|\alpha| = 1$) for $k_a$ non-zero, and is the zero spinor only, when
$k_a = 0$.  We pull back our previous constructions along this
surjection.  The pullback of the ring $K^+$ is then the ring of
functions $f(k_A, k_{A'})$ which are everywhere smooth and vanish to
all orders at the origin, decay faster than any power at infinity and
which obey the differential equation $(k_A \del_k^A -
k_{A'}\del_k^{A'})f = 0$.  We shall need to multiply by components of
the spinors $k_A$ and $k_{A'}$, so it is natural to enlarge the ring
$K^+$ in the spinor case to the ring $\Khat$ which is by definition the
ring of all functions $f(k_A, k_{A'})$, such that $f$ has a
decomposition as an infinite sum: $f = \sum_{r = -\infty}^{\infty}
f_r$, with only a finite number of the functions $f_r$ non-zero and
such that for each integer $r$ we have:
\begin{itemize}
\item[(i)] $f_r$ is globally defined and smooth on the momentum spin
space;
\item[(ii)] $\lim_{t \to \infty} f_r(e^{t\alpha} k_A, e^{t\alpha'}
k_{A'}) = 0$, for all $\alpha \in \CC$, such that $\Re(\alpha)$ is
non-zero; here the limit is taken with $t$ real and the limit must be
uniform on compact subsets of the momentum spin space;
\item[(iii)] $(k_A\del_k^A - k_{A'} \del_k^{A'})f_r(k_A, k_{A'}) =
rf_r(k_A, k_{A'}).$
\end{itemize}
For each integer $s$, denote by $K^s$ the subspace of $\Khat$ consisting
of all $f\in \Khat$ with $f_r$ vanishing for all $r$ different from $s$.
Then the pullback of the ring $K^+$ is the ring $K^0$ and one has $K^p
K^q$ included in $K^{p+q}$, for all integers $p$ and $q$.  In
particular $K^0$ is a subring of $\Khat$ and every space $K^j$ is a
$K^0$-module.  Denote by $K^j[x]$ the subspace of $\Khat[x]$ consisting
of all polynomials in $x$ with coefficients in the space $K^j$.
Denote by $\Lhat[x]$ the subspace of $\Khat[x]$ annihilated by the wave
operator $\dal$ and by $L^j[x]$ the intersection of the spaces
$K^j[x]$ and $\Lhat[x]$.  Denote by $\Khat[\pi]$ the space of all
polynomials in the spinor variables $\pi_A$ and $\pi_{A'}$ with
coefficients in the ring $\Khat$.  For $p$ and $q$ any non-negative
integers and for $j$ any integer, denote by $\Khat_{p,q}[\pi]$ and
$K^j_{p,q}[\pi]$ the spaces of all polynomials in the spinor variables
$\pi_A$ and $\pi_{A'}$, homogeneous of degrees $(p, q)$ in the pair
$(\pi_A, \pi_{A'})$, with coefficients taken from the spaces $\Khat$ and
$K^j$, respectively. For every element $f(x)$ of the space $\Lhat[x]$
and $g(\pi)$ of the space $\Khat[\pi]$, we have unique expansions of the
following form:
\begin{eqnarray}
f(x) &=& \sum_{r = 0}^\infty f_{A_1A_2\ldots A_rA'_1A'_2\ldots A'_r}
x^{A_1A'_1} x^{A_2A'_2}\ldots x^{A_rA'_r},\label{4.1}\\
g(\pi)& = &\sum_{p,q = 0}^\infty
g_{A_1A_2\ldots A_pA'_1A'_2\ldots A'_q} \pi^{A_1}\pi^{A_2}\ldots
\pi^{A_p}\pi^{A'_1}\pi^{A'_2}\ldots \pi^{A'_q}. \label{4.2}
\end{eqnarray}
In equations (\ref{4.1}) and (\ref{4.2}), the coefficient spinors
$f_{A_1A_2\ldots A_rA'_1A'_2\ldots A'_r}$ and $g_{A_1A_2\ldots
A_pA'_1A'_2\ldots A'_q}$ lie in the space $\Khat$ and are completely
symmetric in all indices.  Denote by $E_x^\pi: \Lhat[x] \to \Khat[\pi] $
the evaluation operator which substitutes the spinor $\pi^A\pi^{A'}$
for $x^a$ in any element of $\Lhat[x]$.  Then it is clear that the map
$E_x^\pi$ is an isomorphism of the space $\Lhat[x]$, with range the
subspace of $\Khat[\pi]$ consisting of all polynomials $g(\pi) \in
\Khat[\pi]$, which obey the differential equation $(\pi_A\del_\pi^A -
\pi_{A'}\del_\pi^{A'})g(\pi) = 0$.

To proceed we need the spinor analogues of our previous technical
results.  First the analogue of the surjectivity of the operator $D$
of proposition (\ref{prop5.1}).
\begin{prop}\label{prop5.8}
The operators  $k_A\del^a, k_{A'}\del^a : \Lhat
\to \Lhat$ and $k_A\del_\pi^A, k_{A'}\del_\pi^{A'} : \Khat \to \Khat$
are surjective.
\end{prop}

\noindent{\it Proof:\/} Using the isomorphism $E_x^\pi$ it is easily
seen that
it is sufficent to prove surjectivity for the operators
$k_A\del_\pi^A$ and $k_{A'}\del_\pi^{A'}$.  Further by formal
conjugation the proof of surjectivity for the operator $k_A\del_\pi^A$
will yield a proof of surjectivity for the operator
$k_{A'}\del_\pi^{A'}$.  So we just need to prove that the operator
$k_A\del_\pi^A$ is surjective, when acting on the space $\Khat[\pi]$.
Using the expansion of equation (\ref{4.2}), we reduce to proving that
given a totally symmetric spinor $g_{B\ldots CB'\ldots C'} \in \Khat$,
there exists a totally symmetric spinor $f_{AB\ldots CB'\ldots C'} \in
\Khat$, such that $k^Af_{AB\ldots CB'\ldots C'} = g_{B\ldots CB'\ldots
C'}$.  By taking components with a fixed primed spinor basis, we
reduce further to the case that the spinor $g_{B\ldots CB'\ldots C'}$
has only unprimed indices.  By contracting throughout with a spinor
variable $\pi^A$, we reduce to solving the differential equation
$k_A\del_\pi^A f = g$, given $g \in
\Khat[\pi]$, such that the solution $f$ lies in the space $\Khat[\pi]$
and both $f$ and $g$ are independent of the variable $\pi_{A'}$.

Let $t^a$ denote a fixed unit timelike vector and put $n_A \equiv t_a
k^{A'}$.  Note that $n_A k^A = t_a k^A k^{A'}$ is always a positive
real number unless $k_A = 0$. Then one has the following decomposition
of the function $g$:
\begin{equation}
g = \sum_{p,q = 0}^{\infty} \, {(-1)^p \over p!q!} (k_A\pi^A)^p
(n_B\pi^B)^q g_{p,q}.\label{4.3} \end{equation}
This decomposition follows from the expression of the spinor $\pi_A$
in terms of the spinor basis $n_A$ and $k_A$: $\pi_A =
(t^ck_Ck_{C'})^{-1} (- k_B\pi^B n_A + n_B\pi^B k_A)$.  Then by the
binomial theorem, we have the following explicit formula for the
quantities $g_{p,q}$:
\begin{equation}
g_{p,q} = {1 \over (t^ck_Ck_{C'})^{p+q}} \left[ (n_A\del_\pi^A)^{p}
(k_B\del_\pi^B)^q g\right]_{\pi=0}.\label{4.4}
\end{equation}
It is clear from equation (\ref{4.4}) that
each coefficient $g_{p,q}$ lies in the ring $\Khat$, so by linearity it
suffices to prove the required result for the case $g =
(k_A\pi^A)^{p}(n_B\pi^B)^q$, with $p$ and $q$ non-negative integers.  But then
we have the following relation:
\begin{equation}
k_A\del_\pi^A \left[ {(k_A\pi^A)^{p}(n_B\pi^B)^{q+1} \over (q + 1)
t^ck_Ck_{C'}}\right] = (k_A\pi^A)^{p}(n_B\pi^B)^q .\label{4.5}
\end{equation}
So $f \equiv ((q + 1)t^ck_Ck_{C'})^{-1}(k_A\pi^A)^{p}(n_B\pi^B)^{q+1}$
provides a solution in this case.  Since it is clear that this
function $f$ belongs to the space $\Khat[\pi]$ and is independent of the
variable $\pi_{A'}$, the proof is complete.  Note that by tracking
homogeneities through the proof we find that if $g$ belongs to the
space $K^j_{p,q}$, then we may take the solution $f$ to lie in the
space $K^{j-1}_{p+1,q}$. \qed

Second we need to analyze the kernel of the pullback of the Fourier
transform operator $\cF$.  This Fourier transform, still called $\cF$,
is defined now as follows, when acting on any $\phi \in \Khat[x]$:
\begin{equation}
\cF(\phi) \equiv \int e^{ik_a x^a} \phi(x^a, k_A, k_{A'})\, \Omega.
\label{4.6}
\end{equation}
Here one has $\Omega \equiv \epsilon^{AB}\epsilon^{A'B'} dk_A dk_B
dk_{A'}dk_{B'}$ and the integral is carried out over all of spin
space.  It is easily shown that the operator $\cF$ maps $K^0$
isomorphically onto $\Gamma^+$ (the range of $\cF$ of section 5.2
acting on the space $K^+$) and annihilates all the spaces $K^j$, for
$j$ non-zero.  Furthermore, acting on the space $K^0[x]$ the operator
$\cF$ agrees with the pullback of our original Fourier transform
operator (restricted to the domain $K^+[x]$), up to a fixed non-zero
multiplicative constant.

\begin{prop} \label{prop5.9}: For each $\phi \in L^0[x]$:
\begin{equation}
\phi \in \ker \cF \iff
\phi(x^a, k_A, k_{A'}) = (\del_k^A + ix^ak_{A'})\phi_A + (\del_k^{A'}
+ ix^ak_{A})\phi_{A'},\label{4.7}
\end{equation}
with $\phi_{A}\in L^1[x]$ and $\phi_{A'} \in L^{-1}[x]$ obeying the
spinor zrm-field equations: $\del^a \phi_A = 0$ and $\del^{a}
\phi_{A'} = 0$.  If $\phi$ has degree at most $m$ in $x$, then
$\phi_{A}$ and $\phi_{A'}$ may be taken to have degree at most $m - 1$
in $x$.
\end{prop}
\noindent{\it Proof:\/} The ``if''-part of this result is a trivial
integration by parts, so we assume that $\cF(\phi)$ vanishes and we
establish the formula of equation (\ref{4.7}) for the function $\phi$.
First if $\phi$ is independent of the variable $x$, then $\cF(\phi) =
0$ entails that $\phi = 0$, so the result holds if we take $\phi_A =
\phi_{A'} = 0$.  So now we assume that the required result is true for
$\phi$ any polynomial of degree at most $m - 1$ and take $\phi$ to
have degree at most $m$.  Then applying the wave operator to the
equation $\cF(\phi) = 0$, we obtain the equation $\cF(k_a\del^a\phi)
= 0$, so by the inductive assumption, we have the relation:
\begin{equation}
 k_a\del^a\phi = (\del_k^A + ix^ak_{A'})\psi_A + (\del_k^{A'} +
ix^ak_{A})\psi_{A'}.\label{4.8}
\end{equation}
Here the quantities $\psi_A \in L^1[x]$ and $\psi_{A'} \in L^{-1}[x]$
are polynomials in $x$ of degree at most $m - 2$, belong to the spaces
$L^1[x]$ and $L^{- 1}[x]$ and obey the field equations $\del^a
\psi_A = 0$ and $\del^{a} \psi_{A'} = 0$, respectively.  Write $\phi
= \alpha + \beta$, where $\alpha$ is homogeneous of degree exactly $m$
and $\beta$ is of degree at most $m - 1$ in the variable $x$.
Similarly decompose the fields $\psi_A$ and $\psi_{A'}$ as $\psi_A = -
i\rho_A + \sigma_A$ and $\psi_{A'} = - i\rho_{A'} + \sigma_{A'}$,
where $\rho_A$ and $\rho_{A'}$ are homogeneous of degree $m - 2$,
whereas $\sigma_A$ and $\sigma_{A'}$ have degree at most $m - 3$.
Then the terms of highest degree in the variable $x$ of equation
(\ref{4.8}) give the following equation:
\begin{equation}
k_a\del^a\alpha = x^ak_{A'}\rho_A + x^ak_{A}\rho_{A'}. \label{4.9}
\end{equation}
Note that by the inductive hypothesis, the functions $\rho_A \in
L^1[x]$ and $\rho_{A'} \in \, L^{- 1}[x]$ obey the zrm
field equations: $\del^a \rho_A = 0$ and $\del^{a} \rho_{A'} = 0$.
Note that the quantities $\sigma$, $\rho$ and $\rho'$ are respectively
of homogeneity $(m,m)$, $(m - 1, m - 2)$ and $(m - 2, m - 1)$ in the
variables $\pi_A$ and $\pi_{A'}$, respectively.  Now put $x^a = \pi^A
\pi^{A'}$ in equation (\ref{4.9}).  We obtain the equation:
\begin{equation}
k_a \del_\pi^A \del_\pi^{A'}\sigma = m (k_{A'}\pi^{A'}\rho +
k_{A}\pi^{A}\rho'). \label{4.10}
\end{equation}
Here we have put $\sigma \equiv E_x^\pi(\alpha) \in
K^{0}_{m,m}[\pi]$, $\rho \equiv E_x^\pi(\pi^A \rho_A) \in
K^{1}_{m-1,m-2}[\pi]$ and $\rho'\equiv E_x^\pi(\pi^{A'}\rho_{A'})\in
K^{- 1}_{m-2,m-1}[\pi]$.  Next write $\rho = m^{-1}k_A\del_\pi^A\tau$
and $\rho' = m^{-1}k_{A'}\del_\pi^{A'}\tau'$, for some $\tau \in
K^{0}_{m,m-2}[\pi]$ and $\tau' \in K^{0}_{m-2,m}[\pi]$.  This we can
do by the surjectivity of the operator $k_A\del_\pi^A$ proved above.
Then equation (\ref{4.10}) may be rewritten as follows:
\begin{equation}
0 = k_a (\del_\pi^A \del_\pi^{A'}\sigma -\pi^{A'}\del_\pi^A\tau -
\pi^{A'}\del_\pi^{A'}\tau'). \label{4.11}
\end{equation}

Now suppose that the quantity $v^a \in \Khat[\pi]$ obeys the equation
$k_av^a = 0$.  We may expand the vector $v^a$ in terms of the spinors
$k^A$, $n^A$ and their conjugates $k^{A'}$ and $n^{A'}$ as follows:
\begin{eqnarray}
v^a = k^Ak^{A'}U + k^An^{A'}V &+& n^Ak^{A'}W + n^An^{A'}X,\\
 U \equiv {(v^an_An_{A'}) \over (t_ck^Ck^{C'})^{2}},&\qquad&
 V \equiv -{(v^an_Ak_{A'}) \over (t_ck^Ck^{C'})^{2}}, \\
 W \equiv -{(v^ak_An_{A'}) \over (t_ck^Ck^{C'})^{2}},&\qquad&
 X \equiv {(v^ak_Ak_{A'}) \over (t_ck^Ck^{C'})^{2}}. \label{4.12}
\end{eqnarray}
It is clear that each of the quantities $U$, $V$, $W$ and $X$ lies in
the space $\Khat[\pi]$.  When we have the relation $v^ak_a = 0$, this
implies that the quantity $X$ vanishes.  This in turn entails that the
quantity $v^a$ may be expressed as $v^a = k^Av^{A'} + k^{A'}v^A$, for
some $v^A \in \Khat[\pi]$ and $v^{A'} \in \Khat[\pi]$: indeed one may
take $v^A = Uk^A + Wn^A$ and $v^{A'} = Vn^{A'}$.  Note that if $v^a
\in \, K^{j}_{p,q}[\pi]$, then by equation (\ref{4.12}) the quantities
$v^A$ and $v^{A'}$ may be taken to lie in the spaces
$K^{j+1}_{p,q}[\pi]$ and $K^{j-1}_{p,q}[\pi]$, respectively.

Applying this result to equation (\ref{4.11}), we obtain
\begin{equation}
 \del_\pi^A \del_\pi^{A'}\sigma = \pi^{A'}\del_\pi^A\tau +
\pi^{A}\del_\pi^{A'}\tau' - m^2k^A\upsilon^{A'} -
m^2k^{A'}\upsilon^{A}, \label{4.13}
\end{equation}
for some $\upsilon^{A} \in K^{1}_{m-1,m-1}[\pi]$
and $\upsilon^{A'} \in K^{-1}_{m-1,m-1}[\pi]$.
Contracting equation (\ref{4.13}) through with the spinors $\pi_A$ and
$\pi_{A'}$ gives the following equation:
\begin{equation}
\sigma = k^A\pi_A\upsilon' + k^{A'}\pi_{A'}\upsilon.\label{4.14}
\end{equation}
Here we have put $\upsilon \equiv \pi_{A}\upsilon^{A} \in
K^{1}_{m,m-1}[\pi]$ and $\upsilon' \equiv \pi_{A'}\upsilon^{A'} \in
K^{-1}_{m-1,m}[\pi]$.  Rewriting equation (\ref{4.14}) in terms of the
variable $x$, we find:
\begin{equation}
 \alpha = x^a(k_A\alpha_{A'} + k_{A'}\alpha_{A}).\label{4.15}
\end{equation}
Here the fields $\alpha_A(x)$ and $\alpha_{A'}(x)$ are determined by
the formulas $E_x^\pi (\alpha^A) = m^{-1}\del_\pi^A \upsilon$ and
$E_x^\pi (\alpha^{A'}) = m^{-1}\del_\pi^{A'} \upsilon'$.  Also we
have $\alpha_{A}(x) \in L^{1}[x]$ and $\alpha_{A'}(x) \in L^{-1}[x]$
and both the spinor fields $\alpha_{A}[x]$ and $\alpha_{A'}[x]$ obey
the zrm field equations and are homogeneous of degree $m-1$
in the variable $x$.  By equation (\ref{4.15}), we have the following
relation, using an integration by parts:
\begin{eqnarray}
0 & = &\cF(\phi) = \cF(\alpha + \beta) = \cF(x^a(k_A\alpha_{A'} +
k_{A'}\alpha_{A}) + \beta) \\
 & = &\cF(i\del_k^{A'}\alpha_{A'} + i\del_k^{A}\alpha_{A} + \beta)
\label{4.16}
\end{eqnarray}
By the inductive hypothesis, we obtain from equation (\ref{4.16}) the
relation:
\begin{equation}
i\del_k^{A'}\alpha_{A'} + i\del_k^{A}\alpha_{A} + \beta = (\del_k^A
+ ix^ak_{A'})\omega_A + (\del_k^{A'} + ix^ak_{A})
\omega_{A'}.\label{4.17}
\end{equation}
Here the fields $\omega_{A} \in L^{1}[x]$ and $\omega_{A'} \in
L^{-1}[x]$ obey the spinor zrm field equations and are
polynomials of degree at most $m-2$ in the variable $x$.  Combining
equations (\ref{4.15}) and (\ref{4.17}), we get:
\begin{eqnarray}
\phi = \alpha + \beta & = & x^a(k_A\alpha_{A'} + k_{A'}\alpha_{A}) +
(\del_k^A + ix^ak_{A'})\omega_A + (\del_k^{A'} +
ix^ak_{A})\omega_{A'} - i(\del_k^{A'}\alpha_{A'} +
\del_k^{A}\alpha_{A}) \\
& = &(\del_k^A + ix^ak_{A'})\phi_A + (\del_k^{A'} +
ix^ak_{A})\phi_{A'}. \label{4.18}
\end{eqnarray}
In equation (\ref{4.18}) we have put $\phi_A \equiv \omega_A -
i\alpha_A$ and $\phi_{A'} \equiv \omega_{A'} - i\alpha_{A}$.  Since it
is clear that the fields $\phi_A $ and $\phi_{A'}$ have all the
requisite properties, we have proved the validity of equation
(\ref{4.7}) for any field $\phi (x) \in L^0 [x]$ of degree at most
$m$ in the variable $x$.  Therefore by induction we have the validity
of equation (\ref{4.7}) in general and the proof is complete. \qed

\subsection{The general solution of $M'\Phi=0$}

If we wish to construct a space-time field from elements of $K^j$ with
$j$ non-zero, we first need to multiply by spinors $k_A$ or $k_{A'}$
as appropriate to map the element to an (indexed) element of $K^0$,
before applying the Fourier transform operator $\cF$.  The result is a
spinor indexed field on space-time.  For example consider the standard
zrm equation $\del^{AA'}\Phi_{AB\ldots CD}(x) = 0$ for a totally
symmetric spinor field $\Phi_{AB\ldots CD}(x)$ of $r$ indices.  Taking
another derivative and contracting, we immediately find that the field
$\Phi_{AB\ldots CD}(x)$ obeys the wave equation $\dal\Phi_{AB\ldots
CD}(x) = 0$.  Therefore by the theorem (\ref{thm5.7}), we may write its general
solution, (after pulling back to the momentum spin space) in the space
$\Gamma^+$ , as follows:
\begin{equation}
 \Phi_{AB\ldots CD}(x) = \int e^{ik_a x^a} \alpha_{AB\ldots CD}(k_E,
k_{E'}) \,\Omega.\label{5.1}
\end{equation}
Here the Fourier coefficients $\alpha_{AB\ldots CD}$ lie in the space
$K^0$. Applying the field equation we get the equation
$k^A\alpha_{AB\ldots CD} = 0$, whence it follows that
$\alpha_{AB\ldots CD}(k_E, k_{E'}) = k_Ak_B\ldots k_Ck_D\phi(k_E,
k_{E'})$, for some function $\phi(k_E, k_{E'}) \in K^{-j}$.  So now
equation (\ref{5.1}) reads as follows:
\begin{equation}
 \Phi_{AB\ldots CD}(x) = \int e^{ik_a x^a} k_Ak_B\ldots
k_Ck_D\phi(k_E, k_{E'}) \,\Omega.   \label{5.2}
\end{equation}
Next we shall derive explicitly the solution by Fourier transform of
the equation $\del^{A(A'}\Phi_{AB\ldots C}^{B'\ldots C')} = 0$ in
the special case of one primed index ($m=1$) and then later generalize
to arbitrary positive $m$.  We know that the field $\Phi_{ABB'}$ lies
in the kernel of the operator $\dal^2$ by proposition (\ref{prop2.1}), so by
theorem (\ref{thm5.7}) it admits a Fourier representation of the
following form:
\begin{equation}
 \Phi(X) = \int e^{ik_a x^a} (a_bx^b + a) \,\Omega. \label{5.3}
\end{equation}
Here the variable $X$ is an abbreviation: $X \equiv (x^a, \pi_A,
\pi_{A'})$ and we have put $\Phi(X) \equiv \Phi_{ABB'}(x) \pi^A \pi^B
\pi^{B'}$.  The Fourier coefficients $a_b$ and $a$ depend on the
spinors $k_A$ and $\pi_A$ and their conjugates, but not on the
variable $x^a$.  Defining the operator $M'\equiv
\pi^{A'}\del_\pi^A\del_a$ the field equation may be written as
$M'\Phi=0$. This operator agrees with the operator $M'$ defined in
section 2 in its action on the spinor indexed coefficients of $\Phi$.
Applying the field equation, we get the following equation:
\begin{equation}
0 = \cF\left(\pi^{B'}\del_\pi^B(ik_b + \del_b)(a_cx^c +
a)\right) = \cF \left((\pi^{B'}\del_\pi^B) (ik_b a_cx^c + ik_b a +
a_b) \right). \label{5.4}
\end{equation}
Using equation (\ref{4.7}) above, we deduce the following equation
from equation (\ref{5.4}):
\begin{equation}
 (\pi^{B'}\del_\pi^B) (ik_b a_cx^c + ik_b a + a_b) = (\del_k^A +
ix^ak_{A'})\alpha_A + (\del_k^{A'} + ix^ak_{A})\alpha_{A'}.
\label{5.5}
\end{equation}
Here the quantities $\alpha_A$ and $\alpha_{A'}$ are independent of
the variable $x$.  Equating the coefficients of $x$ in equation
(\ref{5.5}), we get
\begin{equation}
 k_{B'}\pi^{B'}k_B\del_\pi^B a_c =
k_{C'}\alpha_C + k_{C}\alpha_{C'}.
\label{5.6}
\end{equation}
Equation (\ref{5.6}) gives immediately the equation $k_B\del_\pi^B
a^ck_c = 0$, which is solved by $a^ck_c = (\pi^{B}k_{B})^2\alpha$, for
some $\alpha$, independent of the variable $\pi_A$.  This gives the
relation: $a_c = k_{B}\pi^{B}\pi_C\beta_{C'} + k_C\gamma_{C'} +
k_{C'}\gamma_{C}$, for some $\beta_{C'}$, $\gamma_{C'}$ and
$\gamma_{C}$, with $\alpha = k^{A'}\beta_{A'}$.  After an integration
by parts applied to equation (\ref{5.3}), the terms involving
$\gamma_{C'}$ and $\gamma_{C}$ may be eliminated, so one may take just
$a_c = k_{B}\pi^{B}\pi_C\beta_{C'}$, without loss of generality.

Equation (\ref{5.5}) now becomes
\begin{eqnarray}
 &&(\pi^{B'}\del_\pi^B) (ik_b k_{D}\pi^{D}\pi_C\beta_{C'}x^c + ik_b a +
k_{D}\pi^{D}\pi_B\beta_{B'}) \nonumber\\
 &=& ik_{B'}\pi^{B'} k_{D}\pi^{D}k_C\beta_{C'}x^c +
ik_{B'}\pi^{B'}k_B\del_\pi^B a + 3 k_{D}\pi^{D}\pi^{B'}\beta_{B'}\nonumber \\
 &=& (\del_k^A + ix^ak_{A'})\alpha_A + (\del_k^{A'} +
ix^ak_{A})\alpha_{A'}.\label{5.7}
\end{eqnarray}
This gives the relation $k_{A'}\alpha_A + k_{A}\alpha_{A'} =
k_{B'}\pi^{B'} k_{D}\pi^{D}k_A\beta_{A'}$.  So we have $\alpha_{A'} =
k_{B'}\pi^{B'} k_{D}\pi^{D}\beta_{A'} + \delta k_{A'}$ and $\alpha_A =
- k_A\delta$ for some $\delta$.  But then the contribution of the
terms involving the quantity $\delta$ to the right hand side of
equation (\ref{5.7}) is just $(\del_k^A k_A - \del_k^{A'}
k_{A'})\delta = (k_A\del_k^A - k_{A'}\del_k^{A'})\delta$, which
vanishes, since by tracking homogeneities we find that $\delta$ lies
in the space $K^0$.  Therefore without loss of generality, we may take
$\delta = 0$.  Then if we put $x^a = 0$ in equation (\ref{5.7}), we
obtain the relation:
\begin{equation}
ik_{B'}\pi^{B'}k_B\del_\pi^B a + 2 k_{D}\pi^{D}\pi^{B'}\beta_{B'}
 = k_{B'}\pi^{B'} k_{D}\pi^{D}\del_k^{A'}\beta_{A'}.
\label{5.8}
\end{equation}
Next write $\beta_{A'} = \beta_{A'B'}\pi^{B'} + \pi_{A'}\beta$, where
$\beta_{A'B'}$ is symmetric and both $\beta_{A'B'}$ and $\beta$ are
independent of the variables $\pi_A$ and $\pi_{A'}$.  Then putting
$\pi_{A'} = k_{A'}$ in equation (\ref{5.8}) gives the relation:
$\beta_{A'B'}k^{A'}k^{B'} = 0$, which entails that $\beta_{A'B'} =
k_{A'}\delta_{B'}$, for some spinor $\delta_{B'}$.  Then $\beta_{A'} =
k_{A'}\delta_{B'}\pi^{B'} + \pi_{A'}\epsilon $, for some scalar
$\epsilon$.  The term in $\beta_{A'}$ proportional to $k_{A'}$ may be
eliminated by an integration by parts, applied to equation
(\ref{5.3}), so we may take without loss of generality: $\beta_{A'} =
\pi_{A'}\epsilon$. Then equation (\ref{5.8}) reduces to the equation:
\begin{equation}
ik_B\del_\pi^B a
 = k_{D}\pi^{D}\pi_{A'}\del_k^{A'}\epsilon.
\label{5.9}
\end{equation}
Next we put $a = a_{AB}\pi^A\pi^B$, for some symmetric spinor $a_{AB}$,
independent of the spinor $\pi_A$.  Putting $\pi_A = k_A$ in
equation (\ref{5.9}) gives the relation: $a_{AB}k^Ak^B = 0$, so $a_{AB} =
k_{A}\epsilon_{B}$, for some spinor $\epsilon_B$.  Also put $\epsilon_B =
\epsilon_{BB'}\pi^{B'}$, where $\epsilon_b$ is independent of the spinors
$\pi_A$ and $\pi_{A'}$.  Then equation (\ref{5.9}) reduces to the following
equation:
\begin{equation}
\epsilon_{BB'}k^B
 = i\del_k^{A'}\epsilon.
\label{5.10}
\end{equation}
Then we have the relation $a_bx^b + a =
k_{D}\pi^{D}(x^b\pi_B\pi_{B'}\epsilon +
\epsilon_{BB'}\pi^{B}\pi^{B'})$.  Summarizing we have found that the general
solution by Fourier transform of the equation $\del^{A(A'}\Phi_{AB}^{B')}
= 0$ is given by the formula:

\begin{equation}
\Phi(X) = \int e^{ik_a x^a} k_{B}\pi^{B}(x^c\pi_C\pi_{C'}\epsilon +
\epsilon_{CC'}\pi^{C}\pi^{C'}) \,\Omega. \label{5.11}
\end{equation}

Here the quantities $\epsilon$ and $\epsilon_{BB'}$ depend only on the
momentum spinors $k_A$ and $k_{A'}$ and are subject to equation
(\ref{5.10}).  Next equation (\ref{5.10}) may be solved by writing
$\epsilon = - i\phi_A k^A$, for some $\phi_A$.  Then one has
$\epsilon^{AA'} = \del_k^{A'}\phi^A - k^A\phi^{A'}$, for some
$\phi^{A'}$.  The quantities $\phi_B$ and $\phi^{A'}$ are freely
specifiable.  Writing out equation (\ref{5.11}) in terms of the
spinors $\phi_A$ and $\phi_{A'}$, we get the following formula:
\begin{equation}
 \Phi(X) = \int e^{ik_a x^a} k_{B}\pi^{B}(- ix^c\pi_C\pi_{C'}\phi_B k^B +
(\del_k^{C'}\phi^C -
k^C\phi^{C'})\pi_{C}\pi_{C'}) \,\Omega.
\label{5.12}
\end{equation}

Finally we may integrate by parts in equation (\ref{5.12}) to eliminate the
derivative term.  This gives the equation:
\begin{eqnarray}
 \Phi(X) &=& \int e^{ik_a x^a} k_{B}\pi^{B}(-
ix^{CC'}\pi_C\pi_{C'}\phi_B k^B +
i x^{C'C}k_C\phi_B\pi^{B}\pi_{C'} -
k^C\pi_{C}\phi^{C'}\pi_{C'}) \,\Omega \nonumber\\
        &=& \int e^{ik_a x^a} (k_{B}\pi^{B})^2(i x^{C'C}\phi_C\pi_{C'}
+ \phi^{C'}\pi_{C'}) \,\Omega. \label{5.13}%
\end{eqnarray}
Note that there is gauge freedom in the pair of Fourier coefficients
$\phi_\alpha \equiv (\phi_A, \phi^{A'})$: the quantity $\epsilon$ of equation
(\ref{5.10}) is unchanged under the transformation $\phi_A \mapsto \phi_A +
k_A\gamma$.  Then the quantity $\epsilon^{AA'}$ is also unchanged provided we
make the transformation $\phi^{A'} \mapsto \phi^{A'} +
\del_k^{A'}\gamma$.  So the complete gauge transformation is
$\phi_\alpha  \mapsto \phi_\alpha + K_\alpha \gamma$, where $K_\alpha$
is the operator  pair: $K_\alpha \equiv (k_A, \del_k^{A'})$.

Note that equation (\ref{5.13}) may be rewritten in the following
compact form:
\begin{equation}
  \Phi(X) = \int e^{ik_a x^a} (k_{B}\pi^{B})^2 Z^\alpha \phi_\alpha
\,\Omega.    \label{5.14}
\end{equation}
Here $Z^\alpha$ is the twistor $Z^\alpha \equiv (ix^a\pi_{A'},
\pi_{A'})$.  This result may be generalized immediately:

\begin{thm} The integral
\begin{equation}
\Phi(X) = \int e^{ik_a x^a} \phi(Z^\alpha, k_{B}\pi^{B})\,\Omega,
\label{5.15}
\end{equation}
is a solution of the equation $M'\Phi=0$ for every
$\phi(Z^\alpha, k_{B}\pi^{B}) \in \Khat[Z,k_A\pi^A]$. Conversely, let
$\Phi \in \Gamma^+[x,\pi,\pib]$ be a polynomial homogeneous of degree
$(p,q)$ in the spinors $(\pi,\pib)$ with $p > q$. If $\Phi$ satisfies
the equation $M'\Phi=0$ then it has the representation (\ref{5.15})
for some $\phi(Z,k_A\pi^A) \in \Khat[Z, k_A\pi^A]$ homogeneous of
degree $(q,p)$ in the variables $(Z,k_A\pi^A)$.
\end{thm}

The restriction $p > q$ is necessary, because the theorem is false
when $p = q$, or when $p < q$, because in each of these cases the
equation $M'\Phi = 0$ possesses a gauge freedom: if $\Phi =
(\pi_{A'}\pi_{A}\del^a)^p \rho$, where $\rho$ is a polynomial
homogeneous of degree $(0, q)$ in the variables $(\pi_A,\pi_{A'})$,
then the equation $\pi^{A'} \del_\pi^{A} \del_a\Phi = 0$ is
automatically obeyed, for arbitrary such functions $\rho$, as is
checked easily , since the operators $\pi_{A'}\pi_{A}\del^a$ and
$M'$ commute and since $\rho$ is annihilated by the operator $M'$.
Because of this gauge freedom, no Fourier transform formula based on
the null cone of momentum space is possible, unless one first fixes
the gauge freedom in some way.

\noindent{\it Proof:\/} By straightforward differentiation we see
immediately that the function $\Phi$ obeys the equation $\pi^{A'}
\del_\pi^{A} \del_a\Phi = 0$.

Conversely we prove next that the general solution of the equation
$M'\Phi = 0$ may be put in the form of
equation (\ref{5.15}).  The proof is by induction on the integer $q$.
First consider the case $q = 0$.  Then the field $\Phi(X)$ is
independent of the variable $\pi_{A'}$, so the field equation
$M' \Phi = 0 $ is equivalent to the
equation $\del_\pi^{A}\del_a \Phi = 0$, which is just the standard
zrm field equation, for a totally symmetric spinor field with $p > 0$
indices.  By equation (\ref{5.2}) above the solution may be written as
follows:
\begin{equation}
\Phi(X) = \cF ((k_B \pi^B)^p \phi).\label{5.16}
\end{equation}
Here the function $\phi$ is independent of the variables $(x^a, \pi_A,
\pi_{A'})$.  Therefore the required result holds in this case, with the
function $f(Z^\alpha)$ independent of the variable $Z^\alpha$.

Next consider the case $q > 0$.  Let $\Phi(X)$ of homogeneity $(p, q)$
satisfy the field equation $M' \Phi = 0 $
and put $\Psi(X) \equiv \del_\pi^{A'}\del_\pi^{A}\del_a \Phi$.
Then $\Psi(X)$ is of homogeneity $(p -1, q - 1)$ so, since $q - 1$ is
non-negative and since $p - 1 > q - 1$, we may use the inductive
hypothesis and write the field $\Psi(X)$ as follows: $\Psi = \cF({ik_a
x^a}(k_{B}\pi^{B})^{p - 1} \psi(Z^\alpha))$.  Define a field $F(X)$ by
the following formula:
\begin{equation}
F(X) = \cF\left((k_{B}\pi^{B})^{p} f(Z^\alpha)\right).\label{5.17}
\end{equation}
We wish to choose the function $f(Z^\alpha)$, such that the field
$F(X)$ is of homogeneity $(p, q)$ and obeys the equation: $\Psi =
\del_\pi^{A'}\del_\pi^{A}\del_a F$. Applying the differential
operator $\del_\pi^{A'}\del_\pi^{A}\del_a$ to equation
(\ref{5.17}), we get
\begin{eqnarray}
&&\del_\pi^{A'}\del_\pi^{A}\del_a F(X) = \cF((\del_\pi^{A'}
\del_\pi^{A} (\del_a + ik_a)) (k_{B}\pi^{B})^{p}  f(Z^\alpha)) \nonumber\\
&=& \cF(\del_\pi^{A'} \del_\pi^{A} \del_a (k_B\pi^B)^p
f(Z^\alpha))  = - p \cF\left((k_B \pi^B)^{p-1} \del_\pi^{A'} k^A
\del_a f(Z^\alpha)\right)\nonumber \\
& = & i\cF((k_B\pi^B)^{p-1}
\del_\pi^{A'} \pi_{A'}k_{A} (\del_Z)^A f(Z^\alpha)) = ip(q +
1)\cF((k_{B}\pi^{B})^{p-1} k_{A}(\del_Z)^A f(Z^\alpha)). \label{5.18}
\end{eqnarray}
Therefore we have a solution to the equation
$(\del_\pi^{A'}\del_\pi^{A}\del_a) F = \Psi$ provided that the
function $f(Z^\alpha) $ obeys the equation $k_{A}(\del_Z)^A
f(Z^\alpha) = (ip(q + 1))^{-1}g(Z^\alpha)$.

Having found the function $F(X)$, we write $\Phi(X) = F(X) + g(X)$,
for some function $g(X)$.  Then the function $g(X)$ must obey both the
equations $(\del_\pi^{A'}\del_\pi^{A}\del_a) g = 0$ and
$\pi^{A'}\del_\pi^{A}\del_a g = 0$.  Combining these equations, we
get the single equation: $\del_\pi^{A}\del_a g = 0$. By
contracting this equation on the left with the operator
$\pi_{B}\del^{BA'}$, we see that the function $g(X)$ lies in the
kernel of the operator $\dal$ whence it admits a representation
analogous to that of equation (\ref{5.2}): $ g(X) = \cF(\gamma)$, for
some function $\gamma(k_A, k_{A'}, \pi_B, \pi_{B'})$.  Then the field
equation $\del_\pi^{A}\del_a g = 0$ gives the equation $k_A
\del_\pi^A \gamma = 0$, so $\gamma = (k_A \pi^A)^p \eta$, for some
function $\eta(k_A, k_{A'}, \pi_{B'})$.  Putting $\phi(Z^\alpha)
\equiv f(Z^\alpha) + \eta$ now gives the desired representation of the
function $\Phi$ and the complete proof follows by induction. \qed

If we apply this theorem to our special case we obtain the explicit
representation given in the following
\begin{cor} For any non-negative integer $m$ the integral
\begin{equation}
\spinor{\psi}(x) = \int e^{ik_a x^a} k^Ak^B\ldots k^{C}
\left\{ \phi_{B'\ldots C'} + \phi_{E(B'\ldots} x^E{}_{C')}
\cdots + \phi_{E\ldots F} x^E{}_{B'}\cdots x^F{}_{C'} \right\} \Omega,
\end{equation}
is a positive frequency solution of the equation
$\del_{A(A'} \psi^{AB\ldots C}_{B'\ldots C')}=0$. Conversely, every
such solution is represented in the above form.
\end{cor}

\section{Twistor solution of the field equations}

Let us first introduce the structures which are relevant for the
purposes of this work. For more details see \cite{PenroseRindlerII}
and references therein. Twistor space $\TT$ is by definition a
four dimensional complex affine space.  Denote by $\VV$ the
underlying vector space of $\TT$.  At any $z \in \TT$, denote by
$\theta(z)$ the natural isomorphism of the tangent space of $\TT$ at
$z$ with the vector space $\VV$ and denote by $\theta$ the
$\VV$-valued one form on $\TT$ whose value at any $z \in \TT$ is
$\theta(z)$.  It is clear that the one form $\theta$ is exact: $\theta
= d\zeta$.  Here the quantity $\zeta$ is a $\VV$ valued function
globally defined on $\TT$.  The function $\zeta$ is unique up to the
transformation: $\zeta \mapsto \zeta + \alpha$, with $\alpha$
constant.  The function $\zeta$ serves as a vector valued global
coordinate for the space $\TT$.

Naturally associated to the affine space $\TT$ is the space $S(\TT)$
which is the space of all two dimensional affine subspaces of the
space $\TT$.  Naturally associated to the vector space $\VV$ is the
space $M(\TT)$ which is the space of all two dimensional subspaces of
the space $\VV$.  There is a natural surjection, $\mu: S(\TT) \to
M(\TT)$, which takes each element of $S(\TT)$ to its tangent space.
The map $\mu$ renders $S(\TT)$ a two dimensional fibre bundle over
$M(\TT)$ with fibre a two dimensional affine space.  The space
$M(\TT)$ is provided with a natural holomorphic conformal structure,
which is such that $x, y \in M(\TT)$ are null related if and only if
the intersection $x \cap y$ is non-trivial. It is isomorphic to
(compactified, complexified) Minkowski space and the space $S(\TT)$
can be regarded in a natural way as the affine (unprimed) spin bundle
over the space $M(\TT)$.
The primed cospin bundle $S'(\TT)$ is by definition the space of all
pairs $(X, z) \in S(\TT)\times \VV$ with $z$ tangent to $X$.  It is a
two dimensional vector bundle over the space $S(\TT)$.  Denote by
$L'(\TT)$ the line bundle $\Omega^2(S'(\TT))$ over $S(\TT)$.  Note
that the restriction of the form $\theta^2$ to any $X \in S(\TT)$
naturally takes values in the line bundle $L'(\TT)$ at $X$, pulled
back to the space $X$.

Let $f$ denote a holomorphic function defined on some domain $U$ in
$\TT$.  Then the form $f\theta^2$ is a holomorphic two form on
$\TT$ with values in $\Omega^2(\VV)$.  For suitable $X \in
S(\TT)$, consider the following contour integral:
\begin{equation}
    \cS(f)(X) \equiv \int_{\gamma(X)}  f\theta^2. \label{6.1}
\end{equation}

Here $\gamma(X)$ is a closed oriented contour of two real dimensions,
which is required to lie in the intersection of the space $X$ with the
domain of the function $f$ and to vary smoothly with $X$.  It is clear
that the quantity $\cS(f)$ represents a holomorphic section of the
line bundle $L'(\TT)$ over its domain of definition, $M(U)$ (this
domain is an open subset of the space of subspaces $X$, for which the
intersection with the open subset $U$ has non-trivial second
homology).  By definition, if the integration contours are regarded as
given, the section $\cS(f)$ is the (unprimed) spinor field associated
to the twistor function $f$.

More generally let $F(z, \zeta)$ denote a holomorphic function on the
tangent bundle of the domain $U$ with $(z,\zeta) \in
U \times \VV$.  Then consider the following contour integral for
$(X,z) \in S'(\TT)$, such that $X \in M(U)$:
\begin{equation}
\cS(F)(X, z) \equiv \int_{\gamma(X)} F(z,\zeta) \theta^2. \label{6.2}
\end{equation}
This defines a function $\cS(F)$ on $S'(\TT)$ taking values in the
line bundle $L'(\TT)$ (pulled back to the space $S'(\TT)$).

\begin{lem}\label{lem6.1}
 The contour integrals of equation (\ref{6.1}) and
(\ref{6.2}) give coordinate independent formulations of solutions of
the zrm equation and the equation $M'\Phi(x^a, \pi^A, \pi_{A'}) =
\pi^{A'} \del_\pi^A \del_a \Phi(x^a, \pi^A, \pi_{A'}) = 0$,
respectively.
\end{lem}
\noindent{\it Proof:\/} We use lower case greek indices for tensors
based on the vector space $\VV$ and introduce the standard
representation $\zeta^\alpha = (\zeta^A, \zeta_{A'})$ of a twistor
$\zeta$ in terms of an unprimed spinor $\zeta^A$ and a primed cospinor
$\zeta_{A'}$.  A point $X$, not at infinity, of the space $S(\TT)$ is
labelled by the pair $(x^a, \pi^A )$.  The two dimensional affine
subspace of $\TT$ corresponding to $X$ is the set of all twistors
$\zeta^\alpha(\rho_{A'})$ of the form :
\begin{equation}
\zeta^\alpha(\rho_{A'}) = (ix^{AA'}\rho_{A'} + \pi^A,\, \rho_{A'}) =
X^{\alpha A'}\rho_{A'} + \Pi^\alpha, \label{6.3}
\end{equation}
for arbitrary cospinors $\rho_{A'}$. Defining $ X^{\alpha
B'}\equiv (ix^{AB'}, \,\delta_{A'}^{B'})$, $\Pi^\alpha\equiv
(\pi^A,\, 0)$ and $X^{\alpha\beta}\equiv X^{\alpha C'}X^{\beta
D'}\epsilon_{C'D'}$, we note that the restriction of the one form
$\theta^\alpha$ to the space $X$ is given by the formula:
$X^*(\theta^\alpha) = X^{\alpha B'}d\rho_{B'}$ and therefore the
restriction of the form $\theta^\alpha\theta^\beta$ to the space $X$
is just $X^*(\theta^\alpha\theta^\beta) = X^{\alpha\beta}d^2\rho$,
with $d^2\rho \equiv (1/2) \epsilon^{A'B'} d\rho_{A'}
d\rho_{B'}$.  Therefore, the field $\cS(f)(X)$ of equation (\ref{6.1})
factorizes as $\cS(f)(x^c,\pi^C)^{\alpha\beta} =
X^{\alpha\beta}\phi(f)(x^c, \pi^C)$, where we have the following
explicit formulas for the function $\phi(f)(x^a,\pi^A)$:
\begin{eqnarray}
\phi(f)(x^a, \pi^A) & \equiv & \int_{\rho(X)}  f(X^{\alpha
B'}\rho_{B'} + \Pi^\alpha) d^2\rho \nonumber\\
 & = &\int_{\rho(X)}  f(ix^{AB'}\rho_{B'} + \pi^A, \rho_{A'}) d^2\rho.
\label{6.4}
\end{eqnarray}
In equation (\ref{6.4}) the two dimensional contour ${\rho(X)}$ lies
in the primed spin space of the variable $\rho_{A'}$ and varies
smoothly with $X$, avoiding the singularities of the integrand.
Differentiating equation (\ref{6.4}), we immediately obtain the zrm
field equations in the form $\del_\pi^A\del_a\phi(f) = 0$.

Similarly, the integral of equation (\ref{6.2}) gives rise to a field
$\Phi(F)(x^a, \pi^A, \pi_{A'})$ given by the following formula:
\begin{eqnarray}
\Phi(F)(x^a, \pi^A, \pi_{A'}) & \equiv & \int_{\rho(X)}  F(X^{\alpha
B'}\pi_{B'};X^{\alpha B'}\rho_{B'} + \Pi^\alpha ) d^2\rho \nonumber\\
& = &\int_{\rho(X)} F(ix^{AB'}\pi_{B'}, \pi_{A'};ix^{AB'}\rho_{B'} +
\pi^A, \rho_{A'}) d^2\rho. \label{6.5}
\end{eqnarray}
Differentiation of equation (\ref{6.5}) immediately gives the field
equation $M'\Phi(F) = 0$, as required. \qed

Note that, depending on the properties of the twistor functions $f$ or
$F$, the fields $\phi(f)$ and $\Phi(F)$ may contain many different
helicities or irreducible spinor parts.

Denote by $\cO(p, q)$ the sheaf of germs of holomorphic sections of rank
$p$ totally symmetric covariant tensors on projective twistor space
$P(\VV)$, taking values in the sheaf $\cO(q)$ (the sheaf of germs of
holomorphic functions $h(\zeta)$ homogeneous of degree $q$ in the
variable $\zeta$). Such a section is described non-projectively by a
tensor with $p$ indices: $f_{\alpha_1\alpha_2\ldots \alpha_p}
(\zeta^\alpha) \theta^{\alpha_1} \otimes \theta^{\alpha_2}\otimes
\ldots \otimes\theta^{\alpha_p}$, such that $f_{\alpha_1\alpha_2\ldots
\alpha_p}$ is completely symmetric, holomorphic, homogeneous of
integral degree $q - p$ in the variable $\zeta^\alpha$ and such that
$0 = \zeta^{\alpha_1}f_{\alpha_1\alpha_2\ldots \alpha_p}
(\zeta^\alpha)$.  Our main result is that the sheaf cohomology group
$H^1(U, \cO(p, 2p - 1))$, for $p$ a positive integer describes the
general analytic solution to our higher spin equations (the spin is $p
+ 1/2$), for suitable domains $U$ in twistor space. This restriction
to analytic solutions is not mandatory: by replacing ordinary
cohomology by C.R. cohomology one can obtain non-analytic solutions
from the twistor theory.  We do not discuss this further here.

\subsection{Twistor description for the spin (3/2) case}

As before we begin with the spin (3/2) case and treat the general case
later. In this case the object of study is the group $H^1(U, \cO(1,
1))$.  We shall use the contour integral description to get at the
results.  Each calculation that we do then corresponds, according to
well established procedures, to an appropriate calculation using sheaf
cohomology as described for example in the books of Penrose and
Rindler.  For $H^1(U, \cO(1, 1))$ we use functions $f_\alpha(\zeta)$,
which are homogeneous of degree zero in the variable $\zeta$.  For a
function $g(\zeta)$, homogeneous of degree zero, the corresponding
spacetime field $g_{BC}$ is spin one and is given by a contour
integral according to the standard formula of Hughston:
\begin{equation}
g_{BC}(x^a) = \int_{\gamma(x)}\,
\del_B\del_C g(ix^{AB'}\rho_{B'},\rho_{A'}) \rho^{C'}d\rho_{C'}.
\label{7.1}
\end{equation}
Here the operator $\del_B$ denotes the partial derivative with respect
to the unprimed spinor part $\zeta^B$ of the twistor variable
$\zeta^\beta$.  Also the one dimensional contour $\gamma(x)$ is
closed, avoids the singularities of the integrand and varies smoothly
with the point $x$.

Next we need the explicit action of the twistor operator
$\zeta^\alpha$ on the field $g_{BC}$.  Multiplication of $g$ by
$\zeta$ gives a function homogeneous of degree one for which the
corresponding field $\zeta g$ is spin (3/2) and is given by the
following formulas:
\begin{eqnarray}
(\zeta g)^\alpha_{BCD}(x^a) & = &\int_{\gamma(x)}\,
\del_B\del_C\del_D (\zeta^\alpha g)(ix^{EF'}\rho_{F'},\rho_{E'})
\rho^{G'}d\rho_{G'} \nonumber\\
& = &\int_{\gamma(x)}\,
\left(3\delta_{B} ^\alpha\del_C\del_{D}
g(ix^{EF'}\rho_{F'},\rho_{E'}) + X^{\alpha
B'}\rho_{B'}\del_B\del_C\del_D
g(ix^{EF'} \rho_{F'}, \rho_{E'}) \right)\rho^{G'}d\rho_{G'}\nonumber \\
& = & 3\delta_{B}^\alpha g_{CD} - iX^{\alpha B'}\del_{B'B} g_{CD}.
\label{7.2}
\end{eqnarray}
In equation (\ref{7.2}), the twistor $\delta_B^\alpha \equiv
(\delta_B{}^A, 0)$.  Also we have used the fact that inside the
twistor integral the operator $\del_{b}$ is represented by the
operator $i\rho_{B'} \del_B$.  Applying these results to the indexed
function $f_\beta(\zeta)$, allowing for the extra index, equations
(\ref{7.1}) and (\ref{7.2}) become
\begin{eqnarray}
\phi_{\beta CD}(x^a) & = &\int_{\gamma(x)}\,
\del_C\del_D f_\beta(ix^{AB'}\rho_{B'},\rho_{A'}) \rho_{C'}d\rho^{C'}.
\label{7.3} \\
(\zeta\phi)^\alpha_{\beta CDE} & = & 3\delta_{C}^\alpha\phi_{\beta DE} -
iX^{\alpha C'}\del_{C'C}\phi_{\beta DE}. \label{7.4}
\end{eqnarray}
If we now impose the condition $\zeta^\alpha f_\alpha = 0$, we see
that the trace over the twistor indices of equation (\ref{7.4}) must
vanish. From the right hand side of equation (\ref{7.4}) this gives
the following condition:
\begin{eqnarray}
0 & = &3\phi_{CDE} - iX^{\alpha C'}\del_{CC'}\phi_{\alpha DE} =
3\phi_{CDE} + x^{AC'}\del_{CC'}\phi_{ADE} - i\del_{A'C}\phi^{A'}_{DE}
\nonumber\\
& = &3\phi_{CDE} - 2\phi_{CDE} + \del_{CC'}(x^{AC'}\phi_{ADE} -
i\phi^{C'}_{DE}) = \psi_{CDE} - 2\epsilon_{C(D}\psi_{E)} -
i\del_{A'C}\psi^{A'}_{DE}, \label{7.5}
\end{eqnarray}
with $\phi_{\alpha BC} = (\phi_{ABC},\,\phi^{A'}_{DE})$, $\phi_{ABC}
= \psi_{ABC} + \epsilon_{A(B}\psi_{C)}$ and $\psi^{A'}_{DE} \equiv
\phi^{A'}_{DE} + ix^{A'C}\phi_{CDE}$, where the spinors
$\phi^{A'}_{DE}$, $\psi_{ABC}$ and $\psi^{A'}_{DE}$ are completely
symmetric.

Equation (\ref{7.5}) gives in particular the equations $\psi_{CDE} =
i\del_{A'(C}\psi^{A'}_{DE)}$ and $\psi_A = (- i/3) \del^{BB'}
\psi_{ABB'}$.  Hence the field $\phi_{ABC}$ is completely determined
given the field $\psi^{A'}_{AB}$. Once the field $\phi_{ABC}$ is
known, the field $\phi^{A'}_{AB}$ can be recovered from the formula:
$\phi^{A'}_{AB}(x^e) = \psi^{A'}_{AB}(x^e) - ix^{A'C}\phi_{CAB}(x^e)$.
So knowledge of the single field $\psi^{A'}_{AB}$ (and its
derivatives) is completely equivalent to knowledge of the original
field $\phi_{\alpha BC}$.  Finally the field equation for the field
$\phi_{\alpha BC}$ is just the standard zrm field equation
$\del^{B'B}\phi_{\alpha BC} = 0$.  This equation immediately implies
(by straightforward differentiation) the equation
$\del^{B(A'}\psi^{B')}_{BC} = 0$ and conversely it is seen easily that
the equation $\del^{B(A'}\psi^{B')}_{BC} = 0$ implies the field
equation $\del^{B'B}\phi_{\alpha BC} = 0$.  So we have established
that the cohomology group $H^1(U, \cO(1,1))$, for appropriate domains
$U$ in projective twistor space is isomorphic to the space of
solutions of the equation $\del^{B(A'}\psi^{B')}_{BC} = 0$, on the
corresponding domain in space-time, with the field $\psi^{B'}_{BC}$
being totally symmetric.  Finally from the definition of the field
$\psi^{A'}_{AB}$ and from equation (\ref{7.3}), we have the following
contour integral expression for the field $\psi^{A'}_{AB}$:
\begin{equation}
\psi^{A'}_{AB}(x^a) = \int_{\gamma(x)}\, \del_A\del_B f^{A'}
(ix^{AB'}\rho_{B'}, \rho_{A'}) + i x^{A'C}\del_A\del_B
f_{C}(ix^{AB'}\rho_{B'},\rho_{A'}) \rho_{C'}d\rho^{C'}.
\label{7.6}
\end{equation}
Contracting equation (\ref{7.6}) through with $\pi^A \pi^B \pi_{A'}$,
and using Cauchy's integral formula to reduce the integral to a one
dimensional integral, we find complete agreement with equation
(\ref{6.5}), where the function $F(z,\zeta) \equiv z^\alpha f_\alpha$
and the field $\Phi(F)(x^a, \pi^A, \pi_{A'})$ is then a fixed constant
multiple of the field $\psi^{A'}_{AB}(x^e)\pi^A \pi^B \pi_{A'}$. So we
have the shown the

\begin{prop}\label{prop6.2}
 There exists an isomorphism between the sheaf
cohomology group $H^1(U,\cO(1,1))$ and the space of holomorphic
solutions of the equation $\del_{AA'} \phi^{AB}_{B'}=0$.
\end{prop}

\subsection{The general spin case}

Consider the interpretation of the twistor cohomology group $H^1(U,
\cO(p, q))$. We assume that the integers $p$ and $s \equiv q - p + 2$
are positive (later we shall restrict further by requiring that $s >
p$).  As discussed above, a representative element is an indexed
function $f_{\alpha_1\alpha_2\ldots \alpha_p}(\zeta)$, which is
completely symmetric, holomorphic, and homogeneous of degree $q - p$
in the twistor variable $\zeta$ and such that $0 =
\zeta^{\alpha_1}f_{\alpha_1\alpha_2\ldots \alpha_p}(\zeta)$.  The
unprimed spinor field corresponding to the function
$f_{\alpha_1\alpha_2\ldots \alpha_p}$ may be given as follows:
\begin{equation}
\phi(Z^\alpha, x^a, \pi^A) \equiv \int_{\gamma(X)} F(Z^\alpha, X^{\alpha
B'}\rho_{B'} + \Pi^\alpha) d^2\rho.
\label{8.1}
\end{equation}
Here we have put $F(Z, \zeta) \equiv Z^{\alpha_1}Z^{\alpha_2}\ldots
Z^{\alpha_p}f_{\alpha_1\alpha_2\ldots \alpha_p}(\zeta)$.  The field
$\phi(Z^\alpha, x^a, \pi^A)$ is a homogenous polynomial of degree $(p,
s)$ in the pair of variables $(Z^\alpha, \pi^A)$, with coefficients
obeying the zrm field equation by lemma (\ref{lem6.1}).  Our first
objective is to obtain a formula for the action of the twistor
variable $\zeta^\beta $ on such a field.  Denote the result of this
action by $(\zeta\phi)^\beta$.  Then for this field we have the
expression: $(\zeta\phi)^\beta(Z, x,
\pi) = X^{\beta C'}\phi_{C'}(Z, x, \pi) + \Pi^\beta \phi$, where the
field $\phi_{B'}$ is given by the following formula:
\begin{equation}
\phi_{C'}(Z, x, \pi) =  \int_{\gamma(X)} \rho_{C'}F(Z^\alpha,
X^{\alpha B'}\rho_{B'} + \Pi^\alpha) d^2\rho.
\label{8.2}
\end{equation}
Multiplying both sides of equation (\ref{8.2}) by $s + 1$, we manipulate
equation (\ref{8.2}) as follows:
\begin{eqnarray}
(s + 1)&&\phi_{C'}(Z, x, \pi) =  \pi_A \del_\pi^A \,
\int_{\gamma(X)} \rho_{C'}F(Z^\alpha, X^{\alpha B'}\rho_{B'} +
\Pi^\alpha)
d^2\rho \nonumber\\
 &&=  \pi^A\int_{\gamma(X)}
\rho_{C'}((\del_\zeta)_A F)(Z^\alpha, X^{\alpha B'}\rho_{B'} +
\Pi^\alpha) d^2\rho  =  - i\pi^A \del_{AC'}\phi(Z, x, \pi).
\label{8.3}
\end{eqnarray}
Summarizing we have established the formula
\begin{equation}
(\zeta\phi)^\beta = \pi^A \left( - i(s + 1)^{-1}X^{\beta
A'}\del_{a} + \delta_A^\beta\right)\phi. \label{8.4}
\end{equation}
Next, we consider the condition $0 = \zeta^{\alpha_1}
f_{\alpha_1\alpha_2\ldots \alpha_p} (\zeta)$, which is equivalent to
the condition $0 = \zeta^\alpha (\del_Z)_\alpha F$. When applied to
equation (\ref{8.4}), with the field $\phi$ replaced by the field
$(\del_Z)_\alpha \phi$ this implies
\begin{equation}
0 = \pi^A \left( iX^{\beta A'}\del_{a} - (s + 1)\delta_A^\beta \right)
(\del_Z)_\beta \phi = - \pi^A x^{BA'}\del_{a}(\del_Z)_B\phi -
(s + 1)\pi^B(\del_Z)_B\phi + i\pi^A\del_{a}\del_Z^{A'}\phi.
\label{8.5}
\end{equation}
Next we perform a change of variables and write the twistor $Z^\alpha
= (Z^A, Z_{A'})$ as follows: $Z^\alpha = X^{\alpha B'}z_{B'} + z^B
\delta_B^\alpha$, so we have $Z^A = i x^a z_{A'} + z^A$ and $Z_{A'} =
z_{A'}$.  Then the field $\phi$ becomes a function of the variables
$(x^a, \pi^A, z^\alpha)$, where $z^\alpha \equiv (z^A, z_{A'})$.
Under this change of variables, we make the derivative replacements:
$\del_a \mapsto \del_a - i z_{A'}(\del_z)_{A}$, $\, \del_\pi^A \mapsto
\del_\pi^A$, $\, (\del_Z)_A \mapsto (\del_z)_A$ and $\, \del_Z^{A'}
\mapsto \del_z^{A'} - ix^a (\del_z)_A$.  Equation (\ref{8.5}) then
becomes
\begin{eqnarray}
0 &=& - \pi^A x^{BA'} \left( \del_a - i z_{A'}\del_{A}\right)
\del_B\phi - (s + 1)\pi^B\del_B\phi + i\pi^A \left( \del_a - i
z_{A'}\del_{A}\right) \left(\del^{A'} - ix^{BA'}\del_B\right) \phi \nonumber\\
 &=&   \pi^A\del_A(- s + 1 + \gamma')\phi +
i\pi^A\del_a \del^{A'}\phi.\label{8.6}
\end{eqnarray}
Here we have put $\del_A \equiv (\del_z)_A$, $\del^{A'}
\equiv(\del_z)^{A'}$ and $\gamma' \equiv z_{A'}\del^{A'}$.  Also
define $\gamma \equiv z^A \del_A$.  Note that the field $\phi$ obeys
the equation $(\gamma + \gamma' - p)\phi = 0$.

Equation (\ref{8.6}) may be regarded as giving a partial propagation
of the field $\phi$ in the $z^A$ directions.  But we also have the
field equation obeyed by $\phi$, which in terms of the original
variables is the zrm equation $\del_\pi^A \del_a \phi = 0$.  In terms of
the new variables this equation becomes the equation:
\begin{equation}
 0 = \del_\pi^A \del_a \phi - iz_{A'}\del_\pi^A\del_{A}\phi.\label{8.7}
\end{equation}
Removing the factor $\pi^A$ from equation (\ref{8.6}), we get the
following equation, valid for some field $\chi$:
\begin{equation}
 \del_A(- s + 1 + \gamma') \phi +
i\del_a \del^{A'}\phi = \pi_A \chi.
\label{8.8}
\end{equation}
Applying the operator $\del_\pi^A$ to (\ref{8.8}), we get
\begin{equation}
 (s + 1)\chi = \del_\pi^A\del_A(- s + 1 + \gamma')
\phi + i\del_\pi^A\del_a \del^{A'}\phi.
\label{8.9}
\end{equation}
Then, applying the operator $i\del^{A'}$ to (\ref{8.7}), gives
\begin{equation}
  0 = (\gamma' + 2)\del_\pi^A \del_A\phi + i\del^{A'}\del_\pi^A
\del_a \phi,
\label{8.10}
\end{equation}
and, finally, subtracting equation (\ref{8.9}) from equation
(\ref{8.10}) yields
\begin{equation}
 \chi = -  \del_\pi^A \del_A\phi.
\label{8.11}
\end{equation}
{}From equations (\ref{8.8}), (\ref{8.10}) and (\ref{8.11}), we get the
following relation:
\begin{equation}
 (2 + \gamma')(s - 1 - \gamma')\del_A
 \phi = i(2 + \gamma')\del_a \del^{A'}\phi - i\pi_A
\del^{B'}\del_\pi^B\del_b\phi,
\label{8.12}
\end{equation}
and we also have a field equation which is obtained from equation
(\ref{8.7}) by contraction with the spinor $z^{A'}$:
\begin{equation}
 0 = z^{B'}\del_\pi^B\del_b \phi.\label{8.13}
\end{equation}
For the present purposes, equations (\ref{8.12}) and (\ref{8.13}) are
the key equations.  Note that equation (\ref{8.6}) is a consequence of
equation (\ref{8.12}), since the operator $ (2 + \gamma')$ is
invertible.  To analyze these equations most easily, we henceforth
restrict to the case that the quantity $s - p$ is positive.  The
quantity $\del_A \phi$ and the right hand side of equation
(\ref{8.12}) are each sums of terms homogeneous of degrees $0$ to $p -
1$ in the variable $z_{A'}$.  So with $s - p$ positive, the operator
$(s - \gamma' - 1)$ has a well defined inverse acting on such
quantities.  So we may rewrite equation (\ref{8.12}) as follows:
\begin{equation}
 \del_A
 \phi = i(s - 1 - \gamma')^{-1}(\del_a \del^{A'}\phi - (2 +
\gamma')^{-1}\pi_A \del^{B'}\del_\pi^B\del_b\phi).
 \label{8.14}
\end{equation}
We need to check the compatibility of equations (\ref{8.13}) and
(\ref{8.14}).  First we show that the $\del_A$ derivative of the
righthand side of equation (\ref{8.13}) vanishes modulo the equations
(\ref{8.13}) and (\ref{8.14}):
\begin{eqnarray*}
\del_A z^{B'}&& \del_\pi^B\del_b \phi =
z^{B'}\del_\pi^B\del_b \del_A \phi
 = i z^{B'}\del_\pi^B\del_b (s - 1 - \gamma')^{-1}(\del_a
\del^{A'}\phi - (2 + \gamma')^{-1}\pi_A
\del^{C'}\del_\pi^C\del_c\phi) \\
&& = i (s - \gamma')^{-1}(1 + \gamma')^{-1}\left( (1 + \gamma')
z^{B'}\del_\pi^B\del_b \del_a
\del^{A'}\phi - z^{B'}\del_\pi^B\del_b\pi_A
\del^{C'}\del_\pi^C\del_c\phi \right) \\
&& = i(s - \gamma')^{-1}(1 + \gamma')^{-1}(- (1 + \gamma')
\del_\pi^B\del_b \del_a \epsilon^{B'A'}\phi - z^{A'}\del_a
\del^{B'}\del_\pi^B\del_b\phi + \pi_A
\epsilon^{B'C'}\del_\pi^B\del_b
\del_\pi^C\del_c\phi) \\
&& = (i/2)(s - \gamma')^{-1}(1 + \gamma')^{-1}(- (1 + \gamma')
(\del_\pi)_A\dal\phi - 2z^{B'}\del_a\
\del^{A'}\del_\pi^B\del_b\phi - 2\gamma'\epsilon^{A'B'}\del_a
\del_\pi^B\del_b\phi) \\
&& = (i/2)(\gamma' - s)^{-1}(
(\del_\pi)_A\dal\phi +
2\del_\pi^B\epsilon^{A'B'}\del_a\del_b\phi) \\
&& = (i/2)(\gamma' - s)^{-1}(
(\del_\pi)_A\dal\phi +
\del_\pi^B\epsilon_{AB}\dal\phi) = 0.
\end{eqnarray*}
Finally we need to show that applying the operator $\del^A$ to the
righthand side of equation (\ref{8.14}) gives zero.  So it is
sufficient to show that the quantity $(2 + \gamma') \del^A \del_a
\del^{A'} \phi + \pi^A\del_A\del^{B'}\del_\pi^B\del_b\phi$ vanishes,
modulo  equations (\ref{8.13}) and (\ref{8.14}).  We see this as
follows:
\begin{eqnarray*}
(2 + \gamma')&&\del^A\del_a\del^{A'}\phi +
\pi^A\del_A\del^{B'}\del_\pi^B\del_b\phi = (3 +
\gamma')\del^A\del_a\del^{A'}\phi +
\del^{B'}\del_\pi^B\del_b\pi^A\del_A\phi \\
&& = \del_a\del^{A'}(2 + \gamma')\del^A\phi +
\del^{B'}\del_\pi^B\del_b\pi^A\del_A\phi \\
&& = i\del^a\del_{A'}(s - 1 - \gamma')^{-1}((2 + \gamma')\del_{AC'}
\del^{C'}\phi - \pi_A \del^{B'}\del_\pi^B\del_b\phi) +
i\del^{B'}\del_\pi^B\del_b(s - 1 - \gamma')^{-1} \pi^A
\del_a\del^{A'}\phi \\
&& = - i(s - 2 - \gamma')^{-1}((\pi^A\del^{A'}\del_a)
(\del^{B'}\del_\pi^B\del_b)\phi -
(\del^{B'}\del_\pi^B\del_b)(\pi^A\del^{A'}\del_a)\phi)
= 0.
\end{eqnarray*}
Thus equations (\ref{8.13}) and (\ref{8.14}) are integrable.  If we
now write $\phi = \sum_{k = 0}^p \phi_k$, with $\gamma\phi_k = k
\phi_k$, for $0 \le k \le p$, we get from equations (\ref{8.13}) and
(\ref{8.14})  give the following equations:
\begin{eqnarray}
 0 & = & z^{B'}\del_\pi^B\del_b \phi_k, \label{8.17} \\
 \del_A \phi_{k+1} = i(s - p + k)^{-1} && (\del_a \del^{A'}\phi_{k} -
(1 + p - k)^{-1}\pi_A \del^{B'}\del_\pi^B\del_b\phi_{k}).
\label{8.18}
\end{eqnarray}
Equation (\ref{8.17}) is valid for $0 \le k \le p$. Equation
(\ref{8.18}) is valid for $0\le k < p$.  Note that equation
(\ref{8.18}) entails a recursive formula for the quantities $\phi_k$:
\begin{equation}
  \phi_{k+1} = i(k + 1)^{- 1}(s - p + k)^{-1}(z^A\del_a
\del^{A'}\phi_{k} - (1 + p - k)^{-1}z^A\pi_A
\del^{B'}\del_\pi^B\del_b\phi_{k}), \qquad \hbox{for $0 \le k < p$.}
 \label{8.19}
\end{equation}
Equation (\ref{8.19}) shows explicitly that the entire field $\phi$ is
uniquely determined by the field $\phi_0$.  The integrability of
equations (\ref{8.13}) and (\ref{8.14}) shows that the system of field
equations (\ref{8.7}) for the field $\phi(x^a, \pi^A, z^\alpha)$ is
completely equivalent to the single equation $z^{B'}\del_\pi^B\del_b
\phi_0 = 0$, for the field $\phi_0(x^a, \pi^A, z_{A'})$, in the case
$s > p$.

Summarizing, we have outlined a proof of the

\begin{thm} For any pair of positive
integers $(p, t)$, there is an isomorphism between the twistor sheaf
cohomology group $H^1(U, \cO(p, 2p + t - 2))$ and the space of
holomorphic solutions of the field equation $z^{B'}\del_\pi^B\del_b
\phi = 0$, where the field $\phi$ is homogeneous of degrees
$(p+t,p)$ in the variables $(\pi^A, z_{A'})$.
\end{thm}
In particular, we have the
\begin{cor} The space of holomorphic solutions of the equations
$\del_{A(A'}\psi^{AB\ldots C}_{B'\ldots C')}=0$ for spinor fields with
$m+1$ unprimed and $m$ primed indices is isomorphic to
the twistor sheaf cohomology group $H^1(U, \cO(m+1,m))$.
\end{cor}

We would like to make several remarks at this point. Firstly, although
we have shown the existence of the twistor correspondence for several
different kinds of fields (i.e., with different index structures when
considered as spinor fields on space-time) it is only the fields with
homegeneity $(p+1,p)$ that can consistently propagate on a curved
manifold. All other fields suffer from the existence of consistency
conditions. The twistor treatment in this work is to some extent new
in that we use an affine twistor space. This allows to incorporate all
homogeneities into one formula (see e.g., equation (\ref{6.4}) in
comparison to (\ref{7.1})).

\subsection{The group representation}

We observe that there are natural operators acting on the twistor
cohomology groups $H^1(U, O(p,q))$.  Indeed consider the operators
$P^{\alpha \beta}$ and $Q_{\alpha \beta}$ and $E_{\beta}^{\alpha}$,
which act on a representative function $F(z, \zeta)$, homogeneous of
degrees $(p,q-p)$ in the twistor variables $(z, \zeta)$ and obeying
the differential equation $\zeta \cdot\del_z F = 0$, as follows:
\begin{eqnarray}
P(F) &=& (z \wedge \zeta)F, \\
Q(F) &=& (\del_z \wedge \del_{\zeta})F, \\
E(F) &=& (z \otimes \del_z + \zeta \otimes \del_{\zeta} +
\delta)F. \label{6.31}
\end{eqnarray}

In equation (\ref{6.31}), the operator $\delta$ is the Kronecker delta
tensor acting on the representative $F$ by multiplication.  Note that
each of these operators commutes with the operator $\zeta\cdot\del_z$.
The operator $P$ gives a map from $H^1(U, O(p,q))$ to $H^1(U,
O(p+1,q+2))$, the operator $Q$ maps $H^1(U, O(p,q))$ to $H^1(U,
O(p-1,q-2))$ and the operator $E$ maps $H^1(U, O(p,q))$ to itself.
These operators generate a Lie algebra under commutation.  Indeed, by
direct calculation, we have the following commutation relations:

\begin{eqnarray}
 [P^{\alpha \beta}, P^{\gamma \delta}] = 0,&\qquad&
 [Q_{\alpha \beta}, Q_{\gamma \delta}] = 0, \nonumber \\ \relax
 [P^{\alpha \beta}, Q_{\gamma \delta}] =
- \delta_{[\gamma}^{[\alpha}E_{\delta]}^{\beta]},&\qquad&
[E_{\delta}^{\gamma}, P^{\alpha \beta}] = - 2\delta_{\delta}^{[\alpha}
P^{\beta]\gamma},  \label{6.32} \\ \relax
[E_{\delta}^{\gamma}, Q_{\alpha \beta}] = 2\delta_{[\alpha}^{\gamma}
Q_{\beta]\delta},&\qquad&
[E_{\delta}^{\gamma}, E_{\beta}^{\alpha}] = \delta_\delta^\alpha
E_\beta^\gamma - \delta_\beta^\gamma E_\delta^\alpha. \nonumber
\end{eqnarray}

A dimension count gives dimension twenty-eight for this algebra, six
for each of the operators $P$ and $Q$ and sixteen for the operator
$E$.  The operator $E$ generates the complex general linear algebra
$GL(4, \CC)$.  The algebra $GL(4,\CC)$ in turn is isomorphic to the
conformal orthogonal algebra $CO(6,\CC)$ (the orthogonal algebra
together with a dilation).  Adding in the operators $P$ and $Q$ to
this algebra gives the complete algebra of $O(8,\CC)$, regarded as the
conformal algebra associated to $O(6,\CC)$, with $P$ forming the
translations, $Q$ the generator of special conformal transformations,
the tracefree part of $E$ generating rotations and the trace of $E$
giving the dilation.  If we introduce the standard pseudo-hermitian
form on twistor space of signature $(2,2)$, then this algebra has the
natural real form $O(4, 4)$, with the operator $iE$ self-conjugate and
$Q$ the pseudo-hermitian conjugate of $P$.  So we have shown that the
direct sum over $p$ of all the cohomology groups $H^1(U, O(p, 2p + t -
2)$, gives, for each fixed $t$, a complex representation of the Lie
algebra of the group $O(4,4)$.  It remains an open question whether or
not this representation is unitarizable.  Indeed the "natural" inner
product, derived from the action of section one above is not positive
definite in the case of spin greater than one half.  So it would seem
that the representation is "naturally" defined on a space with a
"natural" inner product, but not a Hilbert space.  If one took such
representations seriously, it would apparently require enlarging the
framework of quantum mechanics to accomodate "negative probabilities".

Finally we note that although this algebra is most easily derived in
the twistor picture, one can easily translate into the spacetime
picture, using the techniques of this section.  In the spacetime
picture the operator $E$ acts on the fields as the (complex) conformal
algebra of spacetime.  The operators $P$ and $Q$ and $E$ act as follows on a
field $\phi(x^a, \pi_{A'}, \pi^A)$ obeying the equation $M'\phi = 0$
and homogeneous of degree $s$ in the spinor $\pi^A$:

\begin{eqnarray}
P^{\alpha \beta}\phi &=& {1\over (s + 1)} \pi_{C'} \pi^{C} X^{C'[\alpha}
P^{\beta]}_C \phi, \label{6.33}\\
Q_{\alpha \beta}\phi &=& -{1\over (s + 1)}(\del_\pi)_C
\del_{\pi}^{C'} X^{C}_{[\alpha} Q_{\beta]C'}\phi, \label{6.34}\\
E_{\beta}^{\alpha}\phi & = & - i X^{D'\alpha}X_{\beta}^{D}\del_d \phi +
X^{D'\alpha}\delta_{C'\beta}\pi_{D'}\del_{\pi}^{C'}\phi - X_\beta^D
\delta_C^\alpha (\del_\pi)_D (\pi^C \phi), \label{6.35}\\
\noalign{where we have used the following definitions:}
P^{\alpha}_B &\equiv& (s + 1) \delta^{\alpha}_B - i X^{B'\alpha}\del_b,
\quad Q_{\alpha B'} \equiv (s + 1)\delta_{\alpha B'} - iX_\alpha^B
\del_b,\\
X^{B'\alpha} &\equiv& (ix^{B'A}, \delta_{A'}^{B'}),\qquad
X_{\alpha}^B \equiv (\delta_A^B, - ix^{BA'}), \qquad
\delta^{\alpha}_B \equiv (\delta_B^A, 0),\qquad \delta_{\alpha B'} = (0,
\delta_{B'}^{A'}).
\end{eqnarray}

\section{Conclusion}

We have presented the properties of a class of linear equations for
fields with half integer spin $m+\half$ which generalize the Weyl
equation for a neutrino. We have shown how the equations arise as
Euler-Lagrange equations for a variational principle. The equations
are of hyperbolic type in the sense that the Cauchy problem is well
posed and that there exists the notion of a domain of influence. The
characteristics of the system are multiply sheeted. The fields
propagate freely on any curved background, i.e., there are no
constraints on spatial hypersurfaces to be satisfied by the Cauchy
data. The solutions lie in certain Gevrey classes provided that the
Cauchy data and the metric of the underlying manifold do so. We find a
strong relationship between the spin of the fields and the smoothness
of the metric, ranging from only $\cC^k$ in the neutrino case up to
analyticity in the limit $m\to\infty$. We analyzed the characteristic
initial value problem using the formal method of exact set and showed
that it is well posed in the curved background case as well as when
the system is coupled to gravity via the Einstein equation. It is
interesting to note, that this is a system of partial differential
equations that is not symmetrically hyperbolic (unless $m=0$) but
still allows the description via an exact set. All other examples of
exact sets so far have been systems of equations which were also
symmetrically hyperbolic. This implies that those two
characterizations are not mutually included one in the other. We have
given the general solution of the equations in Minkowski space by
first solving the equation $\dal^{m+1}\phi=0$ using Fourier methods
and then deriving the Fourier representation for positive frequency
fields. Finally, we presented a twistor correspondence between the
the space of holomorphic solutions and sheaf cohomology groups on
projective twistor space.

The solution space of the equations in flat space is a representation
space of the Poincar\'e group. In contrast to the case of the massless
free fields, however, this representation is reducible unless $m=0$.
This can be easily seen from the fact that the entire solution space
for spin $2m-1$ is mapped injectively into the solution space for spin
$2m+1$ by the operator $L$. The solution space is also a
representation space for the conformal group.  It is not yet known
whether this representation is irreducible.

It would be interesting to find similar classes of consistent higher
spin equations for integer spin generalizing the Maxwell
equations. So far attempts have been unsuccesful.

\section*{Acknowledgments}

This work has partly been supported by the SCIENCE program of the
Commission of the European Communities. We would like to thank H.
Friedrich for pointing us towards reference \cite{LerayOhya-1970}.

\appendix

\section{Gevrey classes of functions}

Essential in the proof of existence and uniqueness of solutions of
non-strictly hyperbolic systems of partial differential equations is the
notion of Gevrey classes of functions. These are sets of
$\cC^\infty$-functions, labelled by a real number $\alpha\ge1$ which
in some way interpolate between analytic functions ($\alpha=1$) and
functions which are only $\cC^k$ (conventionally made to
correspond to the case $\alpha=\infty$, see below).
\begin{defn}
Let $S$ be an open set in $\RR^l$, $p\ge1$ and
$\alpha\ge1$. Then $\gamma^{(\alpha)}_p(S)$ is the set of functions
$f: S\to \CC$ such that
\begin{equation}
\sup_\sigma {1\over(1+|\sigma|)^\alpha}\norm{D^\sigma
f,S}_p^{1\over|\sigma|}  < \infty ,
\end{equation}
where $\sigma$ is a multi-index $\sigma=(\sigma_1,\ldots,\sigma_l)$,
$|\sigma|= \sigma_1+\ldots+\sigma_l$ and $\norm{f,S}_p$ is the usual
$L^p$-norm of $f$.
\end{defn}
Similar classes are defined to characterise the behaviour of the
functions with time.

\begin{defn}
Let $\Sigma:=[0,T]\times S$ be a strip in $\RR^{l+1}$,
$p\ge1$, $n\ge1$ and $\alpha\ge1$. Then
$\gamma^{n,(\alpha)}_p(\Sigma)$ is the set of all functions $f: \Sigma
\to \CC$ such that
\begin{equation}
\sup_{\sigma,\beta,x^0} {1\over(1+|\sigma|)^\alpha}\norm{D^{\sigma+\beta}
f,S_t}_p^{1\over|\sigma|}  < \infty ,
\end{equation}
where, again, $\sigma$ and $\beta$ are multi indices with $\sigma_0=0$
(i.e., $\sigma$ refers only to ``spatial'' derivatives) and $0\le
x^0\le T$. $S_t$ is the slice $x^0=t$, as usual.
\end{defn}
We extend the definition with $\alpha=\infty$ by the rule that
$(1+|\sigma|)^\alpha=1$ for $|\sigma|=0$ and $(1+|\sigma|)^\alpha=0$
otherwise. Then $\gamma^{(\infty)}_p(S)$ is equal to the function
space $L_p(S)$ and $\gamma^{n,(\infty)}_p(\Sigma)$ is equal to the set
of all functions for which $\norm{D^\beta f,S_t}$ is a bounded
function of $t$ for all $\beta$ with $|\beta|\le n$.

For $p=\infty$, $l=1$ and for real valued functions this is the
classical case of Gevrey $\cite{Gevrey-1917}$. In that case
$\gamma^{(\alpha)}_\infty(S)$ is an algebra which is closed under the
composition of its elements. A similar property holds in the general
case (see \cite{LerayWahlbroeck-1970}).

Here are some basic properties of the Gevrey classes: they grow with
$\alpha$; if $\alpha_1 \le\alpha_2$ then
$\gamma^{n,(\alpha_1)}_p(\Sigma) \subset
\gamma^{n,(\alpha_2)}_p(\Sigma)$.  Also $\gamma^{n,(\alpha)}_p(\Sigma)
\subset \gamma^{m,(\alpha)}_p(\Sigma)$ if $m\ge n$. If $\alpha=1$ the
classes consist of functions which are analytic in $x^1,\ldots,x^l$,
but for $\alpha\ne1$ one can show that there exists a partition of
unity into elements of the classes with arbitrarily small support;
functions with compact support are not necessarily zero.

The essential qualitative distinction within the Gevrey classes seems
to be between the case $\alpha=1$ and $\alpha>1$, the latter case
permitting domains of influence and thus allowing the study of wave
propagation. This indicates the hyperbolic character of the equations
under consideration. The second distinction is between the cases of
finite $\alpha$ and $\alpha=\infty$.  Infinite $\alpha$ permits the
existence of only a finite number of derivatives and thus the
appearance of shocks is possible indicating the strictly hyperbolic
case.

\def\abstract#1{\hfil\break Abstract: #1} \def\JF{J{\"o}rg Frauendiener}
\ifx\undefined\bysame
\newcommand{\bysame}{\leavevmode\hbox to3em{\hrulefill}\,}
\fi

\end{document}